\documentclass[reprint,amsmath,amssymb,aps,prb,groupedaddress,nofootinbib,twocolumn,superscriptaddress]{revtex4-1}
\usepackage{graphicx}
\usepackage{amsthm,amssymb,amsmath,braket,mathdots}
\usepackage{bm}
\usepackage[pagebackref=false,pdfnewwindow=true]{hyperref} %\hypersetup{draft}
\usepackage{epstopdf,psfrag}
\usepackage{relsize,amsbsy}
\usepackage[export]{adjustbox}

\usepackage{graphicx,xcolor,tikz}

\newcommand{\be}{\begin{equation}}
\newcommand{\ee}{\end{equation}}

\newcommand{\hflux}{heavy flux}
\newcommand{\sbc}{\color{black}{}}

\newcommand{\bit}{\begin{enumerate}}
	\newcommand{\eit}{\end{enumerate}}

\definecolor{bananayellow}{rgb}{1.0, 0.88, 0.21}
\definecolor{straw}{rgb}{0.32, 0.28, 0.1}

\begin{document}

	\title{Dynamics of a quantum spin liquid beyond integrability --\\ the Kitaev-Heisenberg-$\Gamma$ model in an augmented  parton mean-field theory}
		\author{Johannes Knolle}
		\affiliation{\small Blackett Laboratory, Imperial College London, London SW7 2AZ, United Kingdom}
		\author{Subhro Bhattacharjee}
		\affiliation{\small International Centre for Theoretical Sciences, Tata Institute of Fundamental Research, Bengaluru 560089, India}
		\author{Roderich Moessner}
		\affiliation{\small Max-Planck-Institut fur Physik komplexer Systeme, Nothnitzer Str. 38, 01187 Dresden, Germany}
		\date{\today}

		\begin{abstract}
		We present an augmented parton mean-field theory which (i) reproduces the {\it exact} ground state, spectrum, and dynamics of the  quantum spin liquid phase of Kitaev's
		honeycomb model; and (ii) is amenable to the inclusion of integrability breaking terms, allowing a perturbation theory from a controlled starting point. Thus, 
	         we exemplarily study dynamical spin correlations of the honeycomb Kitaev quantum spin liquid within the $K-J-\Gamma$ model which includes 
	         Heisenberg and symmetric-anisotropic {\sbc{(pseudo-dipolar)}} interactions. This allows us to trace  changes of the correlations in the regime of slowly moving fluxes,
	         where the theory captures the dominant deviations when integrability is lost. These include an asymmetric shift together with a broadening of the
	         dominant peak in the response as a function of frequency; the generation of further-neighbour correlations and their structure in real- and spin-space;
	         and a resulting loss of an approximate rotational symmetry of the structure factor in reciprocal space. We discuss the limitations of this approach, and
	         also view the  neutron scattering experiments  on the putative proximate quantum spin liquid material, $\alpha$-RuCl$_3$, in the light of the results from this 
	         extended parton theory.     
		\end{abstract}
	
		\maketitle

\section{Introduction} 
The ground states of systems with long-ranged entanglement and  topological order are notoriously featureless to conventional (local) experimental probes, a fact which greatly complicates their experimental discovery. At the same time, their fractionalised excitations can be much more characteristic and at the same time accessible to regular experiments like inelastic neutron scattering\cite{PhysRevLett.111.137205} (INS), so that they form a natural target in the search for new and exotic states of matter. Such excitations can be thermally excited at finite temperature, but they are also visible even at zero temperature by considering {\it dynamical} correlations in the ground state.

The task of calculating such correlators, however, can be quite {\sbc{complicated}}: while the topological states themselves are usually not easy to describe, capturing their excitations adds another layer of computational complexity as they are highly non-trivial and non-local collective modes in terms of the microscopic degrees of freedom. The way out is often either an exact numerical study on finite-size systems, or the use of approximation schemes whose controllability may be somewhat intransparent.

Against this background, the existence of exactly soluble models has played an important role. While these are few and far between, they have allowed the development of an understanding on a level of detail not available otherwise. This in turn leads to the natural question which aspects of their behaviour are generic, and which are owed to their exact solubility.

The work reported here picks up several strands of this ensemble of questions in the context of the physics of quantum spin liquids  (QSL) which  serve as examples of long-ranged entangled phases of condensed matter, with many experimentally relevant candidate materials~\cite{anderson1987resonating,wen2004quantum, misguich2005frustrated,Lee2008,Balents2010}. Our starting point is the Kitaev honeycomb spin\cite{Kitaev2006} model with its
well-established $Z_2$ quantum spin liquid phase, in which spins fractionalise,
giving way to new degrees of freedom in the form of emergent Majorana
Fermions and $Z_2$ gauge fluxes~\cite{Kitaev2006}. Integrability obtains thanks to the
non-dynamical nature of the latter giving rise to a block-diagonal form of the Hamiltonian, with each
block soluble as a quadratic fermion hopping problem.

This fractionalisation can be captured in a theory of a kind commonly encountered in parton constructions of strongly correlated electron systems in which the spins are written in terms of multiple  fractionalised fermionic and/or bosonic partons which then are expected to capture the nature of the low energy  theory.\cite{PhysRevB.37.580, baskaran1987resonating, PhysRevB.37.3774, PhysRevB.38.5142, RevModPhys.78.17, PhysRevB.65.165113, wen2004quantum, PhysRevLett.62.1694, PhysRevB.42.4568, PhysRevLett.66.1773, arovas1988functional, auerbach1988spin} Such low energy fractionalised excitations transform under  projective representations of symmetries of the system \cite{PhysRevB.65.165113, wen2004quantum, PhysRevB.94.195120}.  Mean field theories (MFTs) of such parton descriptions can then be used to calculate various experimentally relevant quantities such as the dynamic spin structure factor\cite{PhysRevB.88.224413, PhysRevLett.100.227201, Punk2014, PhysRevB.88.174405}--regularly measured in neutron scattering experiments.
 
Indeed, various aspects of the of the Kitaev model have  been understood
in this framework~\cite{Burnell2011,PhysRevB.86.085145, Schaffer2012}. However, only ground state properties are 
exactly reproduced but excited states involving multiple flux excitations are only approximate. 
The underlying reason is that the commonly applied static mean-field approximation removes all dynamic feedback between the different types of emergent fractionalized excitations. 
For example, flux-flux interactions of the Kitaev QSL mediated via the Majorana Fermions are neglected. Thus, such static parton MFTs give strikingly wrong results, even qualitatively, for the dynamic spin structure factor for the Kitaev model (see below).

Our first main result is to extend this parton framework to capture the exact solution of the full excitation 
spectrum at the integrable Kitaev point which requires a more elaborate set-up than the 
previous static approaches which retains a dynamic feedback between the two types of 
excitations~\cite{Punk2014}. 

\begin{figure}
	\centering
	\includegraphics[width=1.1\linewidth]{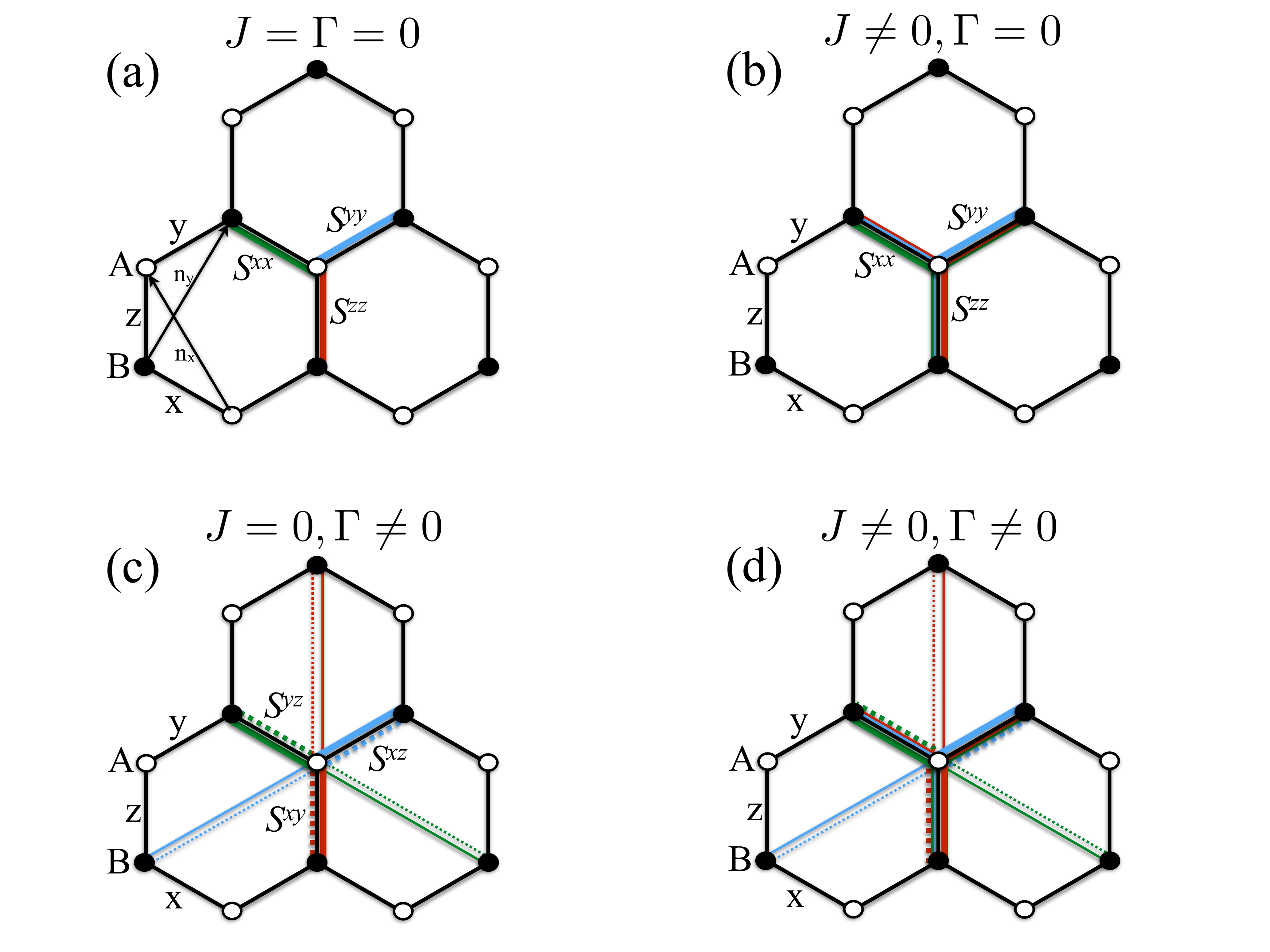}
		\caption{(Color online) The static correlations of the Kitaev-Heisenberg-$\Gamma$ model. Colours denote spin components and line thickness schematically the strength of correlations. In the pure Kitaev model, panel (a), these are ultra-short ranged and bond-selective, e.g. along a nearest neighbour bond $\langle ij\rangle_{\alpha}$ only correlations $S^{\alpha\beta}=\langle S^{\alpha}_i S^{\beta}_j \rangle$ with $\alpha=\beta$ are nonzero and all longer range correlations vanish. With weak Heisenberg interactions, panel (b), the bond selectivity is relaxed but spin correlator remain spin-diagonal and exponentially decaying with distance. A $\Gamma$ perturbation leads exponetially decaying correlations but also to spin-off-diagonal components, panel (c). Both Heisenberg and $\Gamma$ terms lead to a combination of either correlations. In our treatment correlations are still decaying exponentially with distance and neglecting the algebraically decaying tail is justifed by its vanishingly small pre-factor from fourth order perturbation theory~\cite{Song2016}.}
	\label{RealspaceCorrelations}
\end{figure}

The Kitaev spin liquid ground state is expected to be perturbatively stable away from the pure Kitaev model because short-ranged four-femion perturbations, which arise from generic bi-linear interaction like Heisenber exchange, are marginally irrelevant in the renormalisation group sense at the free Majorana fixed point\cite{Burnell2011,Schaffer2012}. These expectations are borne out in recent numerical calculations~\cite{PhysRevLett.105.027204,Gotfryd2017,Gohlke2017,Winter2017} which indicate a smooth evolution of the spin 
dynamics from the Kitaev point throughout the entire QSL phase. 
These results serve as the motivation to study perturbations, which break the integrability of the starting model, within an augmented dynamical parton MFT set-up.

The second {\sbc{important result of}} our work is hence to work out the effect of the removal of the
fine-tuned static nature of the $Z_2$ fluxes responsible for integrability, which manifests itself in
the reappearance of matrix elements/scattering processes which are
absent in the pure Kitaev model. Here, we provide a detailed study
of the changes to the dynamical structure factor, the main ingredient
of which is the addition of dynamics to the gapped flux degrees of freedom. 
This we cannot treat exactly, but instead consider them as a slow variable in a 
\hflux\ approximation, which presents a heuristically motivated first pass at the problem
as long as the fluxes remain gapped and their bandwidth continues to be small compared to the bare energy
scale of the magnetic exchange.

The dynamical structure factor of the pure Kitaev model is dominated by a narrow peak carrying most 
of the total spectral weight above a small gap~\cite{Knolle2014,Knolle2015}.  
In agreement with prior work~\cite{Song2016}, we find that the gap is filled in beyond the static flux limit. 
The most visible consequence of the integrability-breaking terms is 
an asymmetric broadening of the peak in
frequency of the dynamical response on a scale set by the
pertubation. In addition, a qualitative change is that 
previously vanishing correlators acquire non-zero
values, which applies both to correlators between different spin
components, and to correlators between spatially separated spins; however,  
the latter do not do so uniformly but rather retain memory of the
two-sublattice structure of the honeycomb lattice~\cite{Song2016}.

In third step, we also discuss limitations of our approach, which fundamentally arise
from its neglect of multiple-scattering processes {\sbc{between the Majoranas and the fluxes}}, which among
other things are ultimately responsible for recombining the
fractionalised degrees of freedom en route to a magnetic ordering
transition out of the spin liquid.

Our final set of results is a contribution to the current discussion
on the relation of the Kitaev model to the physics of the magnetic
material $\alpha$-RuCl$_3$, which has been billed a proximate spin liquid on
account of its unusual high-energy/finite-temperature dynamical
response~\cite{Banerjee2016,Banerjee2017,Banerjee2017b,Nasu2016,Do2017}, 
for recent reviews see Ref.~\onlinecite{Hermanns2018,Winter2017b}. 
We show that some of the discrepancies between the exact
solution of the Kitaev spin liquid and the experimental results are
remedied as perturbations are added. For example the hexagonal
star-like features of the neutron scattering response appears naturally in this 
treatment and the spectral weight is less concentrated in a single narrow low energy peak.

The remainder of this paper is structured as follows. Sec.~\ref{sec:intrody}
%\ref{sec:apt} 
starts with the definition of the model and introduces notation.  Sec.~\ref{sec:apt}  contains the 
development of the dynamical mean-field theory for the partons, with some technical content relegated to an appendix, which 
the reader not primarily interested in those technical developments may want to skip upon first reading. 
Sec.~\ref{sec:results} contains a detailed  discussion of the properties of the dynamical structure factor summarised above, 
as the central observable of interest in spectroscopic experiments
such as neutron scattering experiments. Its last subsection is devoted to a discussion of the experimental situation in $\alpha$-RuCl$_3$ in the light of
our present theory. We close with a discussion and an outlook in Sec.~\ref{sec:disc}.
%%%%%%%%%%%%%%%%%%%%%%%%%%%%%%%%%%%%

\section{The spin system and the dynamic spin structure factor}
\label{sec:intrody}
Our objective is to study the evolution of the dynamical spin correlations in the Kitaev QSL phase when perturbed from the exactly solvable point. Experimentally, in candidate spin liquid materials, the most interesting set of perturbations to the Kitaev model involve the usual Heisenberg perturbations, and the pseudo-dipolar perturbations. Hence we wish to understand the nature of spin correlations in presence of such perturbations. Thus, our starting point is the so called Kitaev-Heisenberg-$\Gamma$ (KH$\Gamma$) model given by the Hamiltonian
\begin{eqnarray}
\label{HKmodel1}
H & \!= \!& \!\sum_{\langle i j \rangle_{\alpha}} \left\lbrace \!K S_i^{\alpha} S_j^{\alpha} + J \sum_{\beta} S_i^{\beta} S_j^{\beta} +
\Gamma \sum_{\bar \beta \neq\beta \neq \alpha} S_i^{\beta} S_j^{\bar \beta} \right\rbrace
\end{eqnarray}
with spin components $\alpha,\beta=x,y,z$ which also label the three inequivalent bond directions $\langle i j \rangle_{\alpha}$ on the honeycomb lattice, see Fig.\ref{RealspaceCorrelations} (a).  {\sbc{$S_i^\alpha=\sigma^\alpha_i/2$ are the spin-1/2 operators with $\sigma_i^\alpha$ being the Pauli matrices $(\alpha=x,y,z)$.}} We concentrate on isotropic couplings throughout and measure all energies in units of $|K|$. As our work is partially motivated by experimental results on $\alpha$-RuCl$_3$ we follow the emerging consensus~\cite{Winter2017} that the leading interaction is a FM Kitaev term, e.g. for concreteness we set $K=-1$ (FM couplings) and mainly concentrate on FM $J\leq0$ and AFM $\Gamma\geq0$.

 The dynamic spin structure factor,
\begin{align}
\label{StructFact}
\mathcal{S}^{\alpha\beta}({\bf q},\omega)\!=\!\frac{1}{N}\!\sum_{ij}\!e^{i{\bf q}\cdot({\bf r}_i-{\bf r}_j)}\!\!\!\int \!\! \text{d}t\text{d}t'e^{-i\omega(t-t')}\langle S^\alpha_{{\bf r}_i}(t)S^\beta_{{\bf r}_j} (t')\rangle,
\end{align}
 experimentally measured through neutron scattering for magnetic insulators yields direct information about the nature of spin-spin correlations present in the system and hence can shed light on the nature of the ground state {\sbc{of the spin system}}. For the pure Kitaev model the dynamic structure factor can be exactly calculated as the $Z_2$ fluxes are immobile \cite{Knolle2014,Knolle2015,Smith2015,Smith2016} which leads to distinct signatures arising from the strictly non-dispersive gapped $Z_2$ fluxes and gapless Majorana fermions.  A central ingredient in the exact calculation is the non-dispersive nature of the flux which then was used to map the calculations into one of a quenched impurity to obtain the exact solutions. In contrast to the previous work~\cite{Knolle2014,Knolle2015} we wish to go beyond the integrable point which then has to incorporate the central issue of dispersing $Z_2$ fluxes, which in turn maps to a problem of a quenched {\it dynamical} impurity. Given the notorious difficulties of time dependent perturbation theory, particularly with fluxes with gauge strings, we will exploit a different route via an extension of parton MFTs taking into account fluctuations of the emergent $Z_2$ gauge field.

%%%%%%%%%%%%%%%%%%%%%%
\section{Parton Theory for the $HK\Gamma$ model}
\label{sec:apt}

Any parton description of the perturbed Kitaev spin liquid must incorporate the exact solution as a quantitatively well defined limiting case where: (a) the $Z_2$ flux excitations are gapped with the right magnitude of the two flux gap, (b) spin-correlations are exactly nearest neighbour, and (c) the correct excitation spectrum which then produces the right distribution of spectral weights in the dynamic spin structure factor for example. These features then carry on in the perturbative regime with some modifications such as the spin-correlations become exponentially decaying~\cite{Mandal2011} [up to a small algebraic contribution~\cite{Song2016} if both $J\neq0$ and $\Gamma\neq0$ at fourth order $O(J^2\Gamma^2)$]. A characteristic feature of this perturbative regime is a natural separation of scale emerges between the 'fast' itinerant Majorana fermions and the `slow' gapped flux excitations which will allow us to use a `\hflux' approximation to the flux dynamics as detailed below.

% In addition, there a several reasons why a transparent semi-phenomenological  (controlled) theory with the Kitaev point as the unperturbed starting point should be possible for small couplings $J,\Gamma$: 
%i) Flux excitations remain gapped; 
%ii) Spin correlations remain short ranged and exponentially decaying with distance ; 
%iii) Due to the vanishing density of states of the Majorana fermions with a Dirac dispersion the induced Majorana-Majorana interactions are RG-irrelevant, hence they remain gapless~\cite{Hermanns2015}; iv) A natural 

%Given the notorious difficulties of time dependent perturbation theory we will exploit a different route via an extension of parton MFTs taking into account fluctuations of the emergent Z$_2$ gauge field. 

Previous works have shown that simple parton MFTs of the Kitaev model recover some of the exact ground state properties~\cite{Burnell2011,Schaffer2012} which is based on the fact that spins `exactly' fractionalize into itinerant Majorana fermions and $Z_2$ fluxes~\cite{Baskaran2007}, and the emergent $Z_2$ gauge field is static at the exactly solvable point. Hence the fluxes can be thought of as a classical background field for the Majorana fermions. 

However, these simple MFTs fail to correctly capture the full physics of the Kitaev model even at the integrable point. In particular, the description of excited states involving flux excitations is wrong because any feedback between fluxes and fermions is neglected within these MFT approaches where the mean field parameters are static and hence lead to wrong matrix elements arising from the zero flux sector which are absent in the exact calculations because the corresponding matrix elements are zero. Here we augment the MFT with fluctuations which are then used to calculate the correct dynamical response of the spins.

We follow the original work of Kitaev~\cite{Kitaev2006} and use a representation of the spin operator in terms of four Majorana fermions
\begin{eqnarray}
\label{SpinMaj}
S_i^{\alpha}=\frac{1}{2} i c_i b_i^{\alpha}.
\end{eqnarray}
with $\{b_i^{\alpha},b_j^{\beta}\}=2\delta_{ij}\delta_{\alpha\beta}$ (similarly for $c_i$).
The key ingredient for the solubility of the pure Kitaev model, $J=\Gamma=0$, is the presence of an extensive number of local conserved quantities (with eigenvalues $\pm1$)
\begin{align}
W_p=\prod_{ i  \in \partial_p} 2S_i^{\beta} =\prod_{\langle ij \rangle_{\alpha} \in \partial_p} i b_i^{\alpha}
b_j^{\alpha} \delta_{\langle ij \rangle_{\alpha}} .
\end{align}
The product includes all spin operators of a hexagon and there spin component $\beta$ is given by the outward pointing bond type. 
These plaquette fluxes permit a block-diagonalization of the Hamiltonian and the overcomplete representation of the spin Hilbert space in terms of the Majorana fermion operators entails a $Z_2$ gauge redundancy\cite{Yao2007,Pedrocchi2011,Zschocke2015}.

It is convenient to split the Hamiltonian, Eq.\ref{HKmodel1}, into Kitaev- and non-Kitaev contributions
\begin{eqnarray}
\label{HKmodel}
H & = & -\frac{K+J}{4} \sum_{\langle i j \rangle_{\alpha}}  (i b_i^{\alpha} b_j^{\alpha}) i c_i c_j - \frac{J}{4} \sum_{\langle i j \rangle_{\alpha}} \sum_{\beta \neq \alpha} (i b_i^{\beta} b_j^{\beta}) i c_i c_j  \nonumber\\ & & - \frac{\Gamma}{4} \sum_{\langle i j \rangle_{\alpha}} \sum_{\beta \neq \alpha} (i b_i^{\beta} b_j^{\bar \beta}) i c_i c_j .
\label{eq_majham}
\end{eqnarray} 
Explicitly, the combination of spin operators in the last term is for example along an $\alpha=z$ bond: $\left\lbrace \beta,\bar\beta \right\rbrace = \left\lbrace x,y\right\rbrace , \left\lbrace y,x\right\rbrace$.

The presence of the $J$ and $\Gamma$ term renders the model unsolvable by making the $Z_2$ fluxes dynamic, {\sbc{{i.e.}, in presence of $J$ and $\Gamma$, the $W_p$s no longer commute with the Hamiltoninan}}. However, it is very important to ask both from the theoretical side as well as in regards to possible experimental candidates -- what features of the fractionalised excitations, as encoded in the dynamic structure factor survive in presence of such perturbation? 

To answer this question, in absence of the exact solution, a natural candidate is an appropriate parton MFT which can then be utilised to calculate the dynamic spin structure factor.
%%%%%%%%%%%%%%%%%%%%%%

\subsection{Static parton MFT for the ground state of the Kitaev model}

The presence of the formal exact solution for the Kitaev model raises the question of testing the validity, in context of the pure Kitaev model, of mean-field approaches to the fractionalised phases within the ambit of the parton MFTs which have been developed over the years to understand the qualitative features of these phases.\cite{PhysRevB.37.580, baskaran1987resonating, PhysRevB.37.3774, PhysRevB.38.5142, RevModPhys.78.17, PhysRevB.65.165113, wen2004quantum, PhysRevLett.62.1694, PhysRevB.42.4568, PhysRevLett.66.1773, arovas1988functional, auerbach1988spin}

Within such formulation, which naturally gives rise to low energy effective lattice gauge theories with dynamic matter and gauge fields, it is found that the Kitaev ground state is reproduced with some control as the gauge fields become frozen at the exactly solvable point. While several formulations exists\cite{Burnell2011,Schaffer2012}, we focus on the Majorana mean field formalism which we shall use for the rest of this work\cite{PhysRevB.86.085145}. Within this formalism, the Hamiltonian in Eq.\ref{eq_majham} is decoupled using the following mean-field channels:

\begin{eqnarray}
\label{eq_mftpara}
\langle i c_ic_j\rangle & = & \chi^c  \\
\langle  i b_i^{\alpha} b_j^{\alpha}\rangle & = & \chi_K^b \\
\langle i b_i^{\beta} b_j^{\beta}\rangle & = & \chi_J^b  \  \text{with} \  \beta \neq \alpha \\
\langle i b_i^{\beta} b_j^{\bar \beta}\rangle & = & \chi_{\Gamma}^b  \  \text{with} \  \bar\beta \neq \beta \neq \alpha
\end{eqnarray}
which captures the Kitaev ground state-- a single linearly dispersing Majorana fermion ($c$) and three gapped non-dispersing Majorana fermions ($b^\alpha$s) with the flat-band of the latter being related to the conservation of $Z_2$ fluxes of the exact solution. 

The strength of the above MFT becomes evident when one considers finite perturbations $J,\Gamma\neq 0$ to the pure Kitaev model as denoted by Eq.\ref{eq_majham}. While the exact solution is no longer avalilable, the parton MFT can be used to qualitatively understand the effect of the perturbation on the spin liquid ground state and even the transition to a proximate magnetically ordered state  as in Ref.\onlinecite{Schaffer2012}.

However, while the above MFT does faithfully capture the ground state, it fails to account for the spin-spin correlations because within the simple decoupling the spin-spin correlations are given by
$$ \langle S^\alpha_{i}(t)\sigma^\beta_{j}(0)\rangle\sim \langle  i b^\alpha_{i,t}b^\beta_{j,0}\rangle\langle ic_{i,t}c_{j,0}\rangle$$
which yields both qualitatively and quantitatively the wrong spin structure factor as it misses the crucial matrix element effects which mixing different $Z_2$ flux sectors\cite{Baskaran2007,Knolle2014,Knolle2015}.

The lacunae of this static parton MFT need to be rectified to gain control over the calculation of the dynamic structure factor. {\sbc{To this end}}, we present an augmented parton theory which contains the minimal feedback, as explained below, to reproduce the exact dynamic structure factor at the exactly solvable point which then forms the basis of its systematic calculations in presence of small perturbations.

\begin{figure*}
\centering
\includegraphics[width=1.0\linewidth]{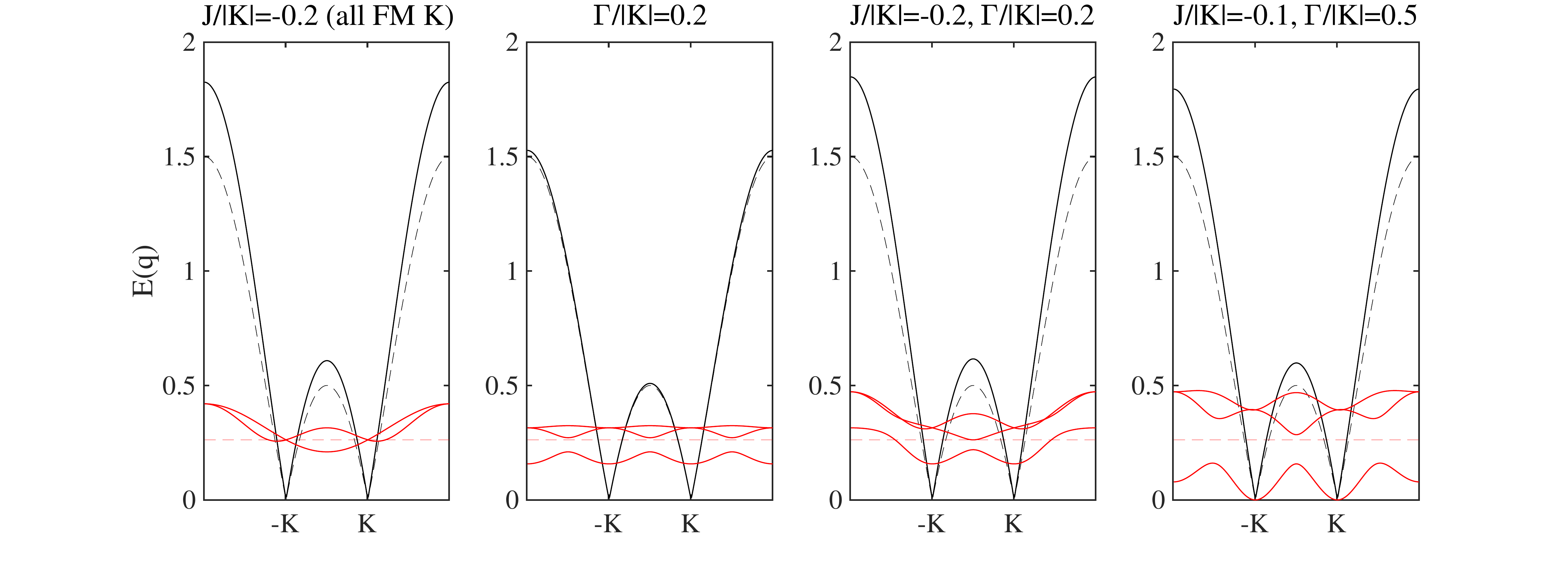}
\caption{(Color online) The MFT dispersions of the matter c-type (black) and flux b-type
(red) Majorana Fermions for different regimes compared to the exactly
soluble point (dashed). The dynamical structure factors for each parameter
set are shown below.}
\label{Energy_Dispersion}
\end{figure*}
\subsection{Augmented Z$_2$ parton MFT}
\label{subsec:z2mft}

We now incorporate the feedback to correct for the dynamics within the
parton description by including fluctuations around the mean-field. 
{\sbc{In addition to the amplitude fluctuations $\hat\delta^{\alpha/c}$ we also include the ones of the overall sign $\hat\sigma^\alpha_{ij}$ corresponding to the $Z_2$ gauge fluctuations. The eigenvalues of $\hat\sigma^\alpha_{ij}$ are $\pm 1$.}} It is motivated by the exact solution in which the link variables $u_{ij}=i
b_i^{\alpha} b_j^{\alpha}$ are constants of motion with eigenvalues $\pm 1$.
Hence, {\sbc{we augment the decoupling channels of Eq.\ref{eq_mftpara}ff as follows.}}  Along an $\alpha$-type
bond $\langle ij \rangle_{\alpha}$
\begin{eqnarray}
\label{MFTparameters}
i c_ic_j & = & \chi^c+\hat\delta_{ij}^{c}  \\
 i b_i^{\alpha} b_j^{\alpha} & = & \hat\sigma_{ij}^{\alpha}
(\chi_K^b+\hat\delta_{ij,K}^{\alpha}) \\
i b_i^{\beta} b_j^{\beta} & = & \chi_J^b+\hat\delta_{ij,J}^{\beta}  \
 \text{with} \  \beta \neq \alpha \\
i b_i^{\beta} b_j^{\bar \beta} & = &
\chi_{\Gamma}^b+\hat\delta_{ij,\Gamma}^{\beta}  \  \text{with} \  \bar\beta
\neq \beta \neq \alpha.
\end{eqnarray}
{\sbc{where we have explicitly put the hat on the fluctuations to point out that they are quantum operators.}} Of particular importance in the rest of this calculation is the  $Z_2$ link
variable with the properties
\begin{eqnarray}
\label{LinkVariable}
\hat\sigma^{\alpha}_{ij}= -\hat\sigma^{\alpha}_{ji} \ \text{and} \ \lbrace
b_i^{\alpha},\hat\sigma_{ij}^{\alpha} \rbrace =0.
\end{eqnarray}
{\sbc{We note that the flavour $\alpha=x,y,z$ of the bond variable $\hat\sigma^\alpha_{ij}$ depends on the type of bond. This}} allows us to minimally incorporate the flux dynamics via the new $Z_2$ link variable which corresponds to the the plaquette
flux of the pure Kitaev model (where $\chi^b_K=1$, see below)
\begin{align}
W_p=\prod_{\langle ij \rangle_{\alpha} \in \partial_p} i b_i^{\alpha}
b_j^{\alpha} \delta_{\langle ij \rangle_{\alpha}} =\prod_{\langle ij
\rangle_{\alpha} \in \partial_p} \hat\sigma_{ij}^{\alpha}.
\end{align}
The spin operators flip the $W_p$'s which is ensured
in our augmented $Z_2$ MFT because we now have the property of the link
variables of the exact solution {\sbc{$\lbrace b_i^{\alpha},u^\alpha_{ij}\rbrace =0$. This introduces a direct coupling between the
$c$-type Majorana matter fermions and the $b$-type flux fermions}}. It is
this ingredient which was missed in previous MFTs and which is crucial for
recovering the full exact solution -- both statics as well as
the dynamics.

%JK: Maybe leave out?
%Note that in principle we should have introduced such $Z_2$ gauge
%fluctuations for $i c_ic_j$, $i b_i^{\beta} b_j^{\beta}$ and $i b_i^{\beta}
%b_j^{\bar \beta}$ as these two link variables are also not gauge
%independent. For $i c_ic_j$, consider a hexagon and the flux of this
%connection through the hexagon. It is given by
%%\begin{align}
%%(ic_1c_2)(ic_3c_2)(ic_3c_4)(ic_5c_4)(ic_5c_6)(ic_1c_6)=+1
%%\end{align} Thus the flux of these through a hexagon, because the majorana
%variables squares to one, is zero. Hence, without loss of generality, the
%corresponding gauge variables are set to be $+1$. In other words fluxes of
%this connection are not an independent degree of freedom. For $i
%b_i^{\beta} b_j^{\beta/\bar\beta}$, this is however more subtle. The
%amplitude of this connection $\chi^b_{J/\Gamma}$ goes to zero when
%$J/\Gamma=0$. Thus these are to be treated as fluctuations which are
%expected to be higher order effects.

% \subsubsection{MFT Hamiltonian}
In a standard MFT approximation we neglect the quadratic part of the amplitude fluctuations, {\sbc{but keeping the phase fluctuations as represented by $\hat \sigma^\alpha_{ij}$}} such that the  Hamiltonian is bilinear and
can be split into matter- and flux-Majorana fermion parts
$H_{\text{MFT}}=H^b+H^c+C$ with
\begin{eqnarray}
 \label{MFTHamiltonian}
 H^c & = & - \frac{1}{4} \sum_{\langle i j \rangle_{\alpha} } \!
\left\lbrace \! \hat\sigma_{ij}^{\alpha} (K\!+\!J) \chi^b_K + 2J \chi^b_J\!
+\!2\Gamma \chi^b_{\Gamma} \! \right\rbrace \! i c_i c_j \\ \nonumber
 H^b & = & - \frac{1}{4} \sum_{\langle i j \rangle_{\alpha} } \left\lbrace
(K+J) \chi^c i b_i^{\alpha} b_j^{\alpha} + J \chi^c \sum_{\beta \neq
\alpha} i b_i^{\beta} b_j^{\beta} \right\rbrace + \\ \nonumber
& & - \frac{1}{4} \sum_{\langle i j \rangle_{\alpha} } \Gamma \chi^c
\sum_{\beta \neq \alpha} i b_i^{\beta} b_j^{\bar\beta}  \\ \nonumber
 C & = & \frac{K+J}{4}\chi^b_K \chi^c \sum_{\langle ij \rangle_{\alpha}}
\hat\sigma_{ij}^{\alpha} + \frac{N_b \chi^c}{2} (J \chi^b_J +\Gamma
\chi^b_{\Gamma})
 \end{eqnarray}
with $N_b$ the total number of bonds. Note, there is an explicit dependence
of the constant $C$ on the configuration of the Z$_2$ variables.

This then forms the augmented parton theory that contains
minimal ingredients required to reproduce the exact results at the Kitaev
point and then can be used to understand the effect of the perturbations on
the dynamic spin structure factor.

 Having set up the augmented parton Z$_2$ MFT, the next steps are
essentially standard derivations of the mean-field solutions including the
enforcement of self-consistency, as detailed in App.~\ref{app:mfsolution}.
From this, we display  the resulting MFT-dispersions, which
demonstrate the (weak) dispersion arising for the previously immobile
fluxes, see Fig.~\ref{Energy_Dispersion}. Note that we have not allowed
for a magnetic ordering instability in this theory, so that the fluxes
become gapless (right panel Fig.~\ref{Energy_Dispersion}) without triggering
a phase transition. We will return to this point in our discussion of
$\alpha$-RuCl$_3$ below.

At this point, we note that previous parton MFTs investigating the HK model including magnetic
decoupling channels found a first order transition from the Kitaev spin
liquid to a gapped fractionalised $U(1)$ spin liquid with simultaneous
magnetic order\cite{Schaffer2012}. It was argued there that the fractionalisation is a
pathological feature of the mean field theory which undergoes confinement
on incorporating instanton effects. Here we do not include the magnetic
parameters and hence our calculations are strictly valid only in the vicinity of
the pure Kitaev limit.

%%%%%%%%%%%%%%%%%

%%%%%%%%%%%%
%%%%%%%%%%%%
\subsection{Dynamical Spin Correlations and Mapping to a Quantum Quench}
Next we focus on our main objective which is to study the dynamical spin correlation function of the Kitaev QSL 
\begin{eqnarray}
\label{SpinCorr}
S_{ij}^{\alpha \beta} (t) & = & \langle 0| S^{\alpha}_i (t) S^{\beta}_j(0)|0\rangle
\end{eqnarray}
and we restrict the discussion to zero temperature {\sbc{when $|0\rangle$ denotes the the ground state}}. 

We rewrite the spin operators in terms of Majorana fermions, Eq.\ref{SpinMaj}, and concentrate on the inter-sublattice component for concreteness
\begin{align}
\label{SpinCorr1}
S^{\alpha \beta}_{A0 B\mathbf{r}} = -\frac{1}{4} \langle 0| e^{it \left(\!H^b\!+\!H^c\!+\!C\!\right)} \!c_{A0} b^{\alpha}_{A0} e^{-it \left(\!H^b\!+\!H^c\!+\!C\!\right)} \! c_{B\mathbf{r}} b^{\beta}_{B\mathbf{r}} |0\rangle.
\end{align}
A major advantage in the {\sbc{augmented}} MFT is the fact that the flux and matter fermions are not directly coupled. Hence, we would like to write the correlator as a simple product of $b$-type and $c$-type correlation functions similar to the derivation for the pure Kitaev model as pioneered by Baskaran et al.~\cite{Baskaran2007}. However, we have to keep in mind that both types of fermions are indirectly coupled via the $Z_2$ link variable. 

First, we split the exponential via the Baker-Hausdorff formula
\begin{eqnarray*}
e^{-it \left(H^b+H^c \right) } = e^{-itH^c} e^{-itH^b} e^{\frac{t^2}{2} \left[H^b,H^c\right]} \approx e^{-itH^c} e^{-itH^b}.
\end{eqnarray*} 
Note, because of our $Z_2$ variable both terms do not commute, but it is easy to show that $\left[H^b,H^c\right]\propto J \chi^b_{J} ...+ \Gamma \chi^b_{\Gamma} ... \propto O(J^2,\Gamma^2)$. Hence, in the limit $J/K,\Gamma/K \ll 1$ ({\rm for~small}~t~{\rm generally}) {\sbc{the commutator}} can be neglected.

Second, we commute the operators past the exponentials. With the commutation properties Eq.\ref{LinkVariable} we obtain 
\begin{eqnarray}
\label{FlipFlux}
b^{\alpha}_{A0}H^c & = & \underbrace{\left[H^c+V^{\alpha}_{A0} \right]}_{ H^c \ \text{with $\sigma^{\alpha}_{A0B\mathbf{n_{\alpha}}}$ flipped}} b^{\alpha}_{A0} \\
b^{\alpha}_{A0} C& = & \left[ C + C^{\alpha}_{A0}\right]  \   \    \  \! \!\!\text{with}  \!\    \  C^{\alpha}_{A0}=-\frac{K+J}{2} \chi^b_K \chi^c.
\end{eqnarray} 
It shows that the action of a spin operator introduces a pair of n.n. fluxes which in turn alters the dynamics of the matter fermions~\cite{Knolle2014,Knolle2015,Knolle2016b}. Hence, the spin correlation function is mapped to the calculation of a quantum quench. {\sbc{From Eq. \ref{SpinCorr1} we thus get}}
\begin{eqnarray}
\label{FlipFlux1}
&&S^{\alpha \beta}_{A0B\mathbf{r}} (t) = \frac{1}{4}e^{iE^c_0 t} \times \\ \nonumber &&\langle 0| c_{A0} e^{-it \left(H^c+V^{\alpha}_{A0} \right)}  e^{-itC^{\alpha}_{A0}}c_{B\mathbf{r}}
e^{itH^b}b^{\alpha}_{A0}e^{-itH^b} b^{\beta}_{B\mathbf{r}} |0\rangle.
\end{eqnarray}
Crucially, the states of the matter sector depend on the flux sector. General states of the MF Hamitonian take the form $|M\rangle=|M_c \left\lbrace \sigma\right\rbrace \rangle |M_b \rangle$ because the occupation of the $b$-type fermions also determine the link variable $\hat\sigma_{ij}^{\alpha}$. For nonzero $J, \Gamma$, fluxes are not static and the flipped flux from Eq.\ref{FlipFlux1} can actually disappear. So, stated differently, the position of the impurity potential seen by the $c$ fermions depends on the dispersion of the $b$ fermions which in turn determine {\sbc{$\langle\hat\sigma_{ij}\rangle=\sigma_{ij}$}}. Hence we can write the dynamical spin correlations as a product of the form
%\begin{widetext}
%\begin{eqnarray}
%S^{\alpha \beta}_{A0B\mathbf{r}} (t) = \delta_{\alpha \beta}\frac{1}{4}e^{iE^c_0 t} \langle 0_c| c_{A0} e^{-it \left[H^c+V^{\alpha}_{A0} \left(\left\lbrace b \right\rbrace \right) \right]} e^{-itC^{\alpha}_{A0}\left(\left\lbrace b \right\rbrace \right)} c_{B\mathbf{r}} |0_c\rangle
%\langle 0_b |e^{itH^b}b^{\alpha}_{A0}e^{-itH^b} b^{\beta}_{B\mathbf{r}} |0_b\rangle.
%\label{eq:fullquench}
%\end{eqnarray}
%\end{widetext}
\begin{eqnarray}
\label{eq:fullquench}
S^{\alpha \beta}_{A0B\mathbf{r}} (t) = \hspace{6.5cm} \\ \nonumber 
\delta_{\alpha \beta}\frac{1}{4}e^{iE^c_0 t} \langle 0_c| c_{A0} e^{-it \left[H^c+V^{\alpha}_{A0} \left(\left\lbrace b \right\rbrace \right) \right]} e^{-itC^{\alpha}_{A0}\left(\left\lbrace b \right\rbrace \right)} c_{B\mathbf{r}} |0_c\rangle \times \\ \nonumber
  \langle 0_b |e^{itH^b}b^{\alpha}_{A0}e^{-itH^b} b^{\beta}_{B\mathbf{r}} |0_b\rangle.
\end{eqnarray}

\subsection{The \hflux\ approximation}
Eq.\ref{eq:fullquench} is in itself not straightforward to evaluate in general. In order to make progress, we need to identify a scheme for evaluating this
expression which is tractable near the exactly soluble point. In the following, we explain such a \hflux\ approximation which, as we discuss further down,
is not systematically controlled but 
correctly captures several of the central features of the change in the dynamical behaviour as integrability is discarded. 

\begin{figure}
	\centering
	\includegraphics[width=1.1\linewidth]{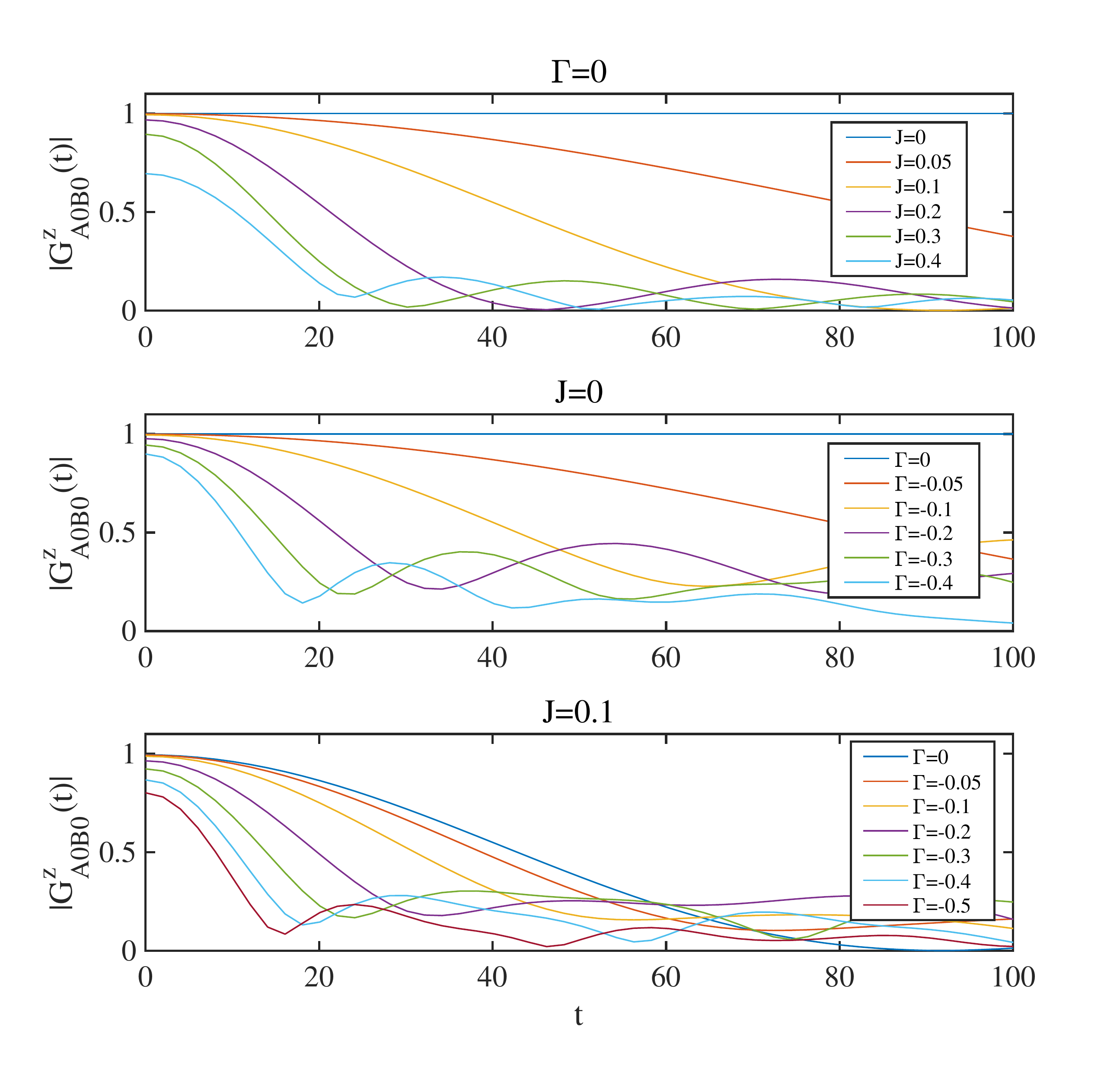}
		\caption{The decay of the flux propagator $G^z_{A0B0}(t)$ which is related to the 
		time dependence of the Z$_2$ link variable is shown for various values of $J,\Gamma$. 
		The average time $\tau$ after which a flux has hopped away can  be read off. Note that the suppression 
		of the propagator below unity at $t=0$ reflects
		the admixture of fluxes into 
		  the ground state, which is flux-free only at the integrable point.}
	\label{Fig4}
\end{figure}

For small $J,\Gamma$ as long as the flux gap remains large the fluxes are much slower (heavier) than the gapless matter fermions. For example, the mass of the fluxes or the $b$ fermions are order $1/J$ for $\Gamma=0$, see the red lines in Fig.~\ref{Energy_Dispersion}.
Looking at the decay time of the flux-Majorana propagator
\begin{eqnarray}
\label{bpropagator}
G^{\alpha}_{A0B\mathbf{r}} (t) = i \langle 0_b | b_{A0}^{\alpha} (t) b^{\alpha}_{B\mathbf{r}}|0\rangle
\end{eqnarray}
at $\mathbf{r}=0$ we can  calculate the average time $\tau$ after which a flux pair has disappeared {\sbc{from a given plaquette by hopping away}}, 
e.g. from the condition $|G^{\alpha}_{A0B0} (t)| \leq \frac{1}{2}$. This directly introduces a time 
dependence for our potential and in the simplest form it is just abruptly switched off once the flux hopped 
away $V^{\alpha}_{A0}(t)=V^{\alpha}_{A0}\Theta(\tau-t)$ {\sbc{(where $\tau$ sets the decay time calculated from the above mentioned condition and $\Theta$ is the Heavyside function)}}. 
The decay of the propagator $|G^{\alpha}_{A0B0} (t)|$ is shown in Fig.\ref{Fig4} for various $J,\Gamma$. 
Note, for small $J$ and $\Gamma=0$ (small $\Gamma$ and $J=0$)  the decay time scales as 
$\tau \propto J^{-1}$ ($\tau \propto \Gamma^{-1}$) as expected from the scaling of the mass of the $b$-fermions.
 %%%%%%%
 
 %%%%%%%%%% 

Finally, with the matter fermion correlator 
\begin{eqnarray}
\label{GcImp}
\!\!\!\!\!\!\!\!G^c_{A0 B\mathbf{r}} (t,\tau) \!= \!i e^{iE^c_0 t} \langle 0_c| c_{A0} e^{-it \left[H^c+V^{\alpha}_{A0} \Theta \left(\tau-t\right) \right]} c_{B\mathbf{r}} |0_c\rangle  
\end{eqnarray} 
we obtain an expression which is a simple product of time dependent matter and flux fermion propagators
\begin{eqnarray}
\label{SpinCorrSemiClassical}
S^{\alpha \beta}_{A0B\mathbf{r}} (t) = -\frac{1}{4}e^{it \Delta_b} G^c_{A0 B\mathbf{r}} (t,\tau) G^{\alpha\beta}_{A0B\mathbf{r}} (t) ,
\end{eqnarray} 
and similarly for $S^{\alpha \beta}_{A0A\mathbf{r}} (t)$ [the explicit definition of $G^{\alpha\beta}_{A0B\mathbf{r}}$ is given below; Note, for $G^c_{A0 A\mathbf{r}} (t,\tau)$ just replace $c_{B\mathbf{r}} \to c_{A\mathbf{r}}$ but the local potential $V^{\alpha}_{A0} $ is always at site $A0$]. This factorized form of the spin correlation function is the main result of the derivation. Crucially, it allows us to calculate the flux and matter Majorana propagator separately via our augmented $Z_2$ MFT.

\begin{figure*}
	\centering
	\includegraphics[width=0.8\linewidth]{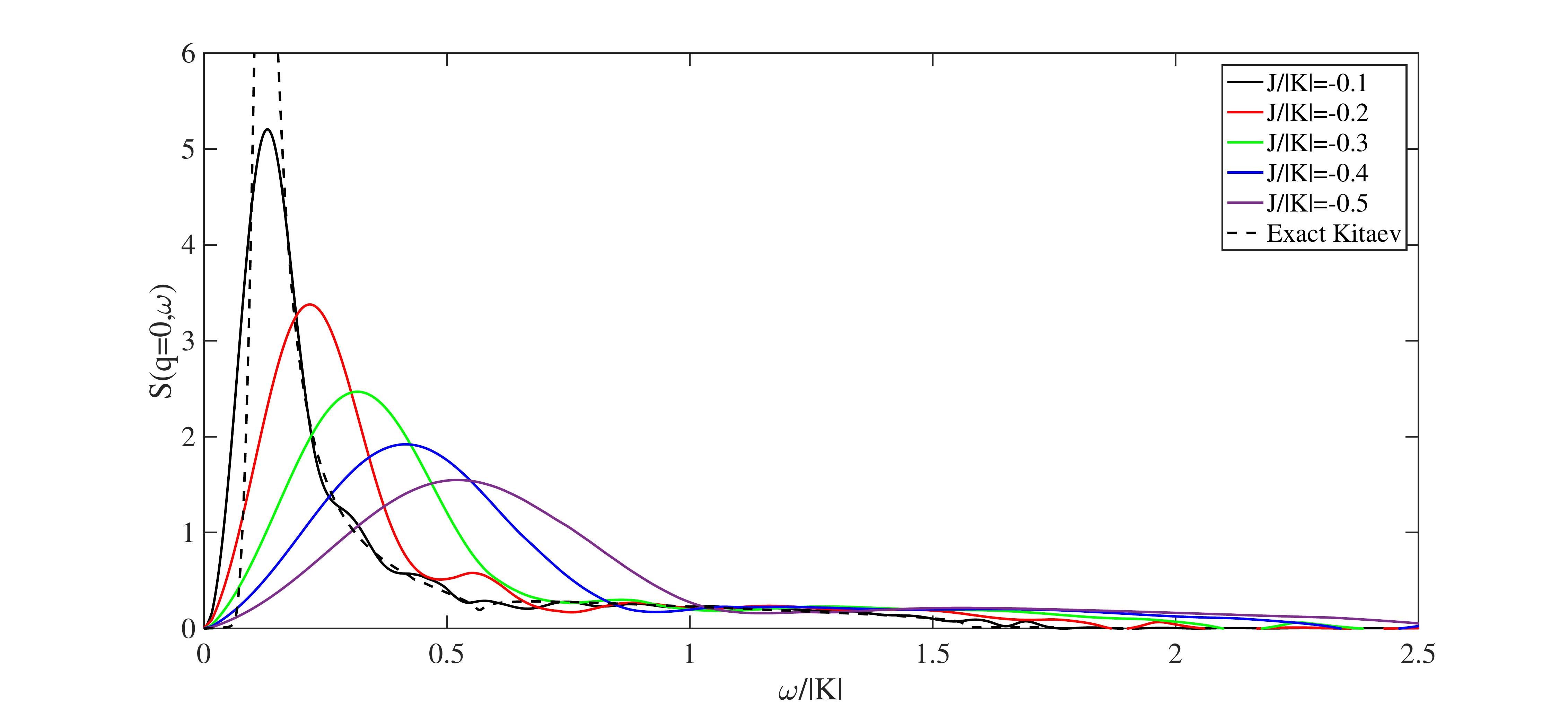}
%		\caption{Evolution of the dominant peak of the structure factor $S(\mathbf{q}=0,\omega)$ of the FM Kitaev model as a function of 
%		perturbing Heisenberg ($J$) (top) and $\Gamma$-interaction (bottom) . Note%
%		the asymmetric broadening accompanying a shift of the centre on a scale set by the strength of the perturbation.}
	\label{Sqzero1}
%\end{figure*}
%\begin{figure*}
%	\centering
	\includegraphics[width=0.8\linewidth]{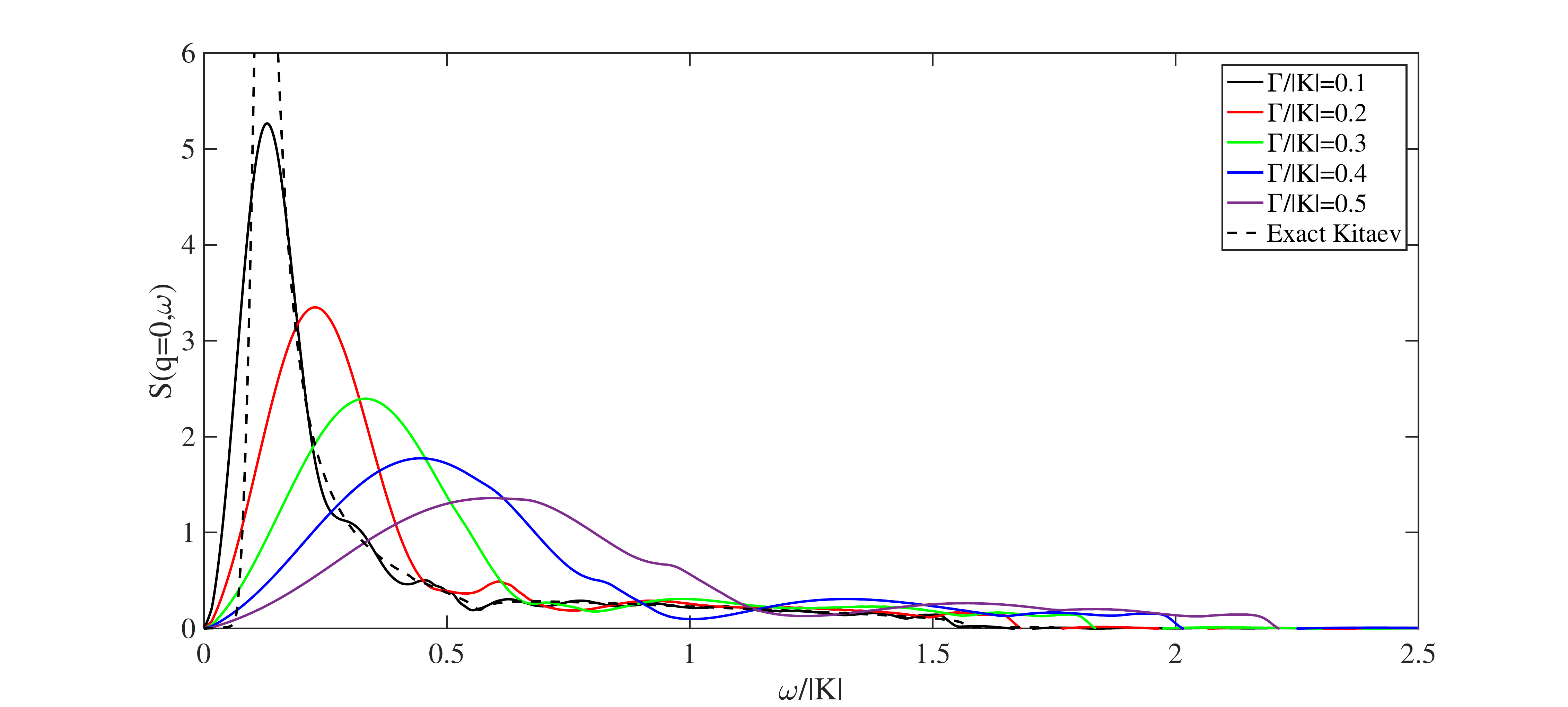}
%		\caption{Evolution of the dominant peak of the structure factor $S(\mathbf{q}=0,\omega)$ of the FM Kitaev model as a function of perturbing $\Gamma$-interaction.}
		\caption{Evolution of the dominant peak of the structure factor $S(\mathbf{q}=0,\omega)$ of the FM Kitaev model as a function of 
		perturbing Heisenberg ($J$) (top) and $\Gamma$-interaction (bottom). Note
		the asymmetric broadening accompanying a shift of the centre on a scale set by the strength of the perturbation.}
	\label{Sqzero2}
\end{figure*}

\subsubsection{Expression for the fermionic propagators} 
For the matter fermions at times $t<\tau$ it is necessary to evaluate the full local quantum quench including the impurity potential, see Eq.~\ref{GcImp}. This is done on large real space lattice ($65\times65$ unit cells) using the methods of our previous works~\cite{Knolle2015,Knolle2016b}. Once the flux has disappeared the calculation is much easier and given by (see App.~\ref{app:mfsolution} for details and definitions)
 \begin{widetext}
 \begin{eqnarray}
G^{c}_{A0B\mathbf{r}} (t>\tau) & = & i \langle 0_c| c_{A0}(t) c_{B\mathbf{r}}|0_c \rangle = \sum_{\mathbf{q}} e^{-it 2|S(\mathbf{q})|} \left\lbrace  e^{i\mathbf{q}  \mathbf{r}} \cos^2 \theta_{\mathbf{q}}  -  e^{-i\mathbf{q} \mathbf{r}} \sin^2 \theta_{\mathbf{q}} -2 \sin \theta_{\mathbf{q}} \cos \theta_{\mathbf{q}}  \sin \mathbf{q} \mathbf{r}\right\rbrace  \\
G^{c}_{A0A\mathbf{r}} (t>\tau) & = &  i \langle 0_c| c_{A0}(t) c_{A\mathbf{r}}|0_c \rangle = i \sum_{\mathbf{q}} e^{-it 2|S(\mathbf{q})|} \left\lbrace  e^{i\mathbf{q}  \mathbf{r}} \cos^2 \theta_{\mathbf{q}}  +  e^{-i\mathbf{q} \mathbf{r}} \sin^2 \theta_{\mathbf{q}} \right\rbrace .
\end{eqnarray}
\end{widetext}
For a smooth Fourier transformation to frequency space we match up the solutions from $G^{c}_{A0B\mathbf{r}} (t<\tau)$ with $G^{c}_{A0B\mathbf{r}} (t>\tau)$. In practice the propagators are glued together as 
\begin{eqnarray}
G^c_{A0B\mathbf{r}}(t,\tau) & = & \left[1-f(t,\tau)\right] G^c_{A0B\mathbf{r}}(t,\tau=t)  \\ \nonumber && +f(t,\tau) G^c_{A0B\mathbf{r}}(t,\tau=0)
\end{eqnarray}
with the smoothed out step function $f(t,\tau)=\frac{1}{1+e^{-t+\tau}}$.

Finally, the flux propagators are calculated from 
 \begin{eqnarray}
G^{\alpha \beta}_{A0B\mathbf{r}} (t) & = & -\sum_{\mathbf{q}} e^{i (\mathbf{q} \mathbf{r}-t\epsilon^j_{\mathbf{q}})} U^{\alpha}_{j \mathbf{q}} V^{\beta}_{j \mathbf{q}} \\
G^{\alpha \beta}_{A0A\mathbf{r}} (t) & = & -i\sum_{\mathbf{q}} e^{i (\mathbf{q} \mathbf{r}-t\epsilon^j_{\mathbf{q}})} U^{\alpha}_{j \mathbf{q}} U^{\beta*}_{j \mathbf{q}} .
\end{eqnarray}
{\sbc{where $U^\alpha_{j\bf q}$ and $V^\alpha_{j\bf q}$ are coherence factors obtained by diagonalising the MFT Hamiltonian as discussed below, see Eq.\ref{HbMft}}} in the appendix.

\subsection{Recovery of the exact solution}
We close this section with the demonstration -- in contrast to previous parton MFT treatements\cite{Burnell2011,Schaffer2012} --
that our augmented MFT (i) recovers the full exact solution including all excited states at the integrable Kitaev point with $J,\Gamma=0$ 
(ii) and as such can be used to compute the exact dynamical correlations at this point as well.

As in the exact solution~\cite{Baskaran2007} the static spin correlations in the ground state are only nonzero for on-site and nearest-neighbor correlators
\begin{eqnarray}
S^{\alpha \beta}_{A0B\mathbf{r}} &=& -\frac{1}{4}  \langle 0| c_{A0} b^{\alpha}_{A0} c_{B\mathbf{r}} b^\beta_{\mathbf{r}}|0\rangle \\ \nonumber
&=&  \delta_{\mathbf{n_\alpha},\mathbf{r}} \delta_{\beta,\alpha} \chi^c \chi^b_K = -\text{sgn} \left[K \right]  0.5249
\end{eqnarray}
because the dispersion of the $b$-type flux Majoranas is a flat band~\cite{Schaffer2012} which also gives $\chi^b_K=1$. 

Excitations of the $c$-type matter fermions in the ground state flux sector are obviously the same as in the exact solution, see Eq.\ref{DiagonalHamiltonian_c}. What about the flux excitations? At first sight it seems that pairs of nearest neighbor flux excitations arise from populating the flat $b$-type bands with energy $E_b=\frac{K\chi^c}{2}$ see Eq.\ref{DiagonalHamiltonian_b}. However, this contradicts the exact solution where a flux excitations costs an energy $E_b= \frac{K\chi^c}{8}$. Moreover, exciting $N$ pairs of fluxes is not just $N$ times $E_b$ from populating $N$ flux b-fermions in the flat band. In the exact solution fluxes interact via coupling to the $c$-type matter Majorana background reducing the overall energy of multiple flux excitations~\cite{Kitaev2006}.

Nevertheless, a careful consideration of the definition of the energy of a flux excitation shows that our augmented MFT is nothing but an alternative form of the exact solution in contrast to previous MFTs~\cite{Burnell2011,Schaffer2012}. The energy of a pair of fluxes is defined as the energy difference
\begin{eqnarray}
E_b=\langle 0|b^z_{A0} H b^z_{A0}|0\rangle -\langle 0| H |0\rangle.
\end{eqnarray}
Keeping in mind all terms in the MFT Hamiltonian Eq.\ref{MFTHamiltonian} including the constants and the fact that the ground state of the matter Fermions depends on the flux background {\sbc{$|0\rangle = |0_c \left\lbrace \sigma \right\rbrace\rangle |0_b\rangle$}} via the link variable $\sigma^{\alpha}_{ij}$ (e.g. $|0_c \left\lbrace +\forall \right\rbrace\rangle$ labels the matter g.s. in the flux free sector with all $\sigma_{ij}=+1$) we see that
\begin{eqnarray*}
\langle 0| H |0\rangle & = & \underbrace{\langle  \left\lbrace +\forall \right\rbrace 0_c| H_c |0_c \left\lbrace +\forall \right\rbrace\rangle}_{E_0^c \left\lbrace + \forall \right\rbrace = -\sum_{\mathbf{q}} |S(\mathbf{q})| }\! +\!\!\! \underbrace{\langle 0_b| H_b |0_b\rangle}_{E_0^b=-\frac{3K\chi^c N_B}{4}} \!\!+ \frac{K \chi^b_K \chi^c N_B}{4} \\ \nonumber
& = & -\frac{K \chi^c N_B}{2}+E_0^c \left\lbrace + \forall \right\rbrace \nonumber
\end{eqnarray*}
and 
\begin{eqnarray*}
\langle 0|b_{A0}^z Hb_{A0}^z |0\rangle & = & \underbrace{\langle  \left\lbrace \! \sigma^z_{A0B0}\!=\!-\!1\! \right\rbrace 0_c| H_c |0_c \left\lbrace \! \sigma^z_{A0B0}\!=\!-\!1\! \right\rbrace\rangle}_{E_0^c \left\lbrace \! \sigma^z_{A0B0}\!=-1\! \right\rbrace} +\\ 
&& \underbrace{\langle 0_b|b^z_{A0} H_b b^z_{A0}|0_b\rangle}_{E_1^b=-\frac{3K\chi^c N_B}{4}+\frac{K\chi^c}{2}} + \frac{K \chi^b_K \chi^c (N_B-2)}{4} \\ \nonumber
&  = & -\frac{K \chi^c N_B}{2}+E_0^c \left\lbrace \sigma^z_{A0B0}=-1  \right\rbrace \nonumber
\end{eqnarray*}
such that the flat band contribution is exactly canceled by the constants and we recover the exact energy of a pair of flux excitations which is nothing but the difference in ground state energies of the matter Majoranas with and without a pair of fluxes in the background
\begin{eqnarray}
E_b=E_0^c \left\lbrace \sigma^z_{A0B0}=-1  \right\rbrace -E_0^c \left\lbrace + \forall \right\rbrace .
\end{eqnarray}
The same reasoning shows that all excited state energies including multiple fluxes are correct in our description. 

It is somewhat surprising that the flat band flux Fermions completely drop out as physical degrees of freedom.

%\subsubsection{Recovery of the exact solution}
With this in hand, we can also demonstrate that the augmented mean-field theory recovers the exact solution of the 
dynamical structure factor. 
Note for $\Gamma=0$ the Hamiltonian is diagonal in the different types of $b^{\alpha}$ Majoranas which directly gives $S^{\alpha \beta} \propto \delta_{\alpha,\beta}$. 
For $J,\Gamma=0$ the $b$-type propagator is simply $G^{z}_{A0B\mathbf{r}}(t) =\delta_{\mathbf{r},0} e^{-it\frac{K\chi^c}{2}}$ which exactly cancels the overall phase, which from the derivation is given by $\Delta_b=\frac{(K+J) \chi^b_K \chi^c}{2}$, in Eq.\ref{SpinCorrSemiClassical} ($K=1,J=0,\chi^b_K=1$). 
The full spin correlator is then determined by the matter Fermion correlator. This is exactly the same expression of a local quantum quench as in the pure Kitaev model~\cite{Knolle2014,Knolle2015}. Thus, our \hflux\ approximation has the correct limit for $J\to 0$ and can be readily compared with our previous work on the exact solution (notation as in Ref.\onlinecite{Knolle2016b})
\begin{eqnarray*}
S^{zz;J,\Gamma=0}_{A0B0} (t) & = & -i \langle 0| e^{iH^c t} c_{A0} e^{-(H^c+V^z_{A0})t} c_{B0} |0\rangle = - G^c_{A0B0} \\
S^{zz;J,\Gamma=0}_{A0A0} (t) & = &  \langle 0| e^{iH^c t} c_{A0} e^{-(H^c+V^z_{A0})t} c_{A0} |0\rangle = -i G^c_{A0A0} .
\end{eqnarray*}

\section{Results}
\label{sec:results}
We concentrate on results obtained for the dynamical structure factor, Eq.\ref{StructFact}, which is the Fourier transform in space and time of  
the dynamical spin correlation function defined in Eq.\ref{SpinCorr}.
Furthermore, to make connection to the inelastic neutron scattering (INS) experiments we also study the intensity of the INS response, $\mathcal{I}(\mathbf{q},\omega)$, which includes the projectors to the transverse momentum components of the spin direction and the overall decay from the form factor $F(|\mathbf{q}|)$:
\begin{eqnarray}
\label{INSInt}
\mathcal{I}(\mathbf{q},\omega) & = & F(q) \sum_{\alpha,\beta} \left(\delta_{\alpha,\beta} -\frac{q^{\alpha} q^{\beta}}{q^2}\right)\mathcal{S^{\alpha \beta}}(\mathbf{q},\omega).
\end{eqnarray}

\subsection{Robust versus fine-tuned features}
Our modified $Z_2$ MFT and approximations for the spin dynamics recover the exact results of the Kitaev point in the limit $J,\Gamma \to 0$. Importantly, it can now be extended beyond the integrable point which allows us to discern the fine-tuned features from the robust qualitative behaviour of the correlations representative of the Kitaev QSL phase.

\subsubsection{Static correlations}
First, we discuss the static correlation which can simply be obtained from our main result Eq.\ref{SpinCorrSemiClassical}
\begin{eqnarray}
\label{SpinCorrStatic}
S^{\alpha \beta}_{A0B\mathbf{r}}= -\frac{\delta_{\alpha\beta}}{4}G^c_{A0 B\mathbf{r}} (t=0)G^{\alpha}_{A0B\mathbf{r}} (t=0).
\end{eqnarray}
Recall, that for the pure Kitaev model correlations are only non-zero for on-site and nearest neighbour correlators. In addition along a $z$-type bond only $S^{zz}$ correlators were non-zero etc. The main qualitative changes induced by Heisenberg and $\Gamma$ interaction are summarised as follows, see also Fig.\ref{RealspaceCorrelations}.

First, correlations between sites of the same sublattice $S^{\alpha \beta}_{A0A\mathbf{r}} = \delta_{0,\mathbf{r}}$ continue to be only on-site. The reason is the bipartite hopping problem of the $c$-type matter Majoranas which is unaffected by the extra couplings. 

Second, with nonzero but small $J$ or $\Gamma$ the real space correlations fall of exponentially because they are governed by the flux-propagators deviating from the flat band limit, which was already predicted in Ref.~\cite{Mandal2011}. Also, with {\it both} nonzero $J,\Gamma$ there is a small fourth order algebraic contribution which we neglect here but which has a negligible influence on the total weight of the structure factor in the perturbative regime around the Kitaev point~\cite{Song2016}. Of course, the phase transition out of the ordered state will be driven by divergence of long range correlations, and therefore, the current augmented MFT is only applicable inside the Kitaev QSL phase. It does not capture its breakdown other than indicating a termination of the stability of the phase through the vanishing of the flux gap {\sbc{whence the heavy-flux approximation clearly breaks down}}. 

Third, the strong bond selectivity of the spin correlations is relaxed beyond the Kitaev limit. A non-zero $J$ induces finite $S^{xx,yy}$ correlations along $z$-bonds (etc.) and a non-zero $\Gamma$ induces finite $S^{xy,yx}$ correlations along $z$-bonds (etc.), see the summary in Fig.\ref{RealspaceCorrelations}. From the quickly decaying static real space spin correlations we can reduce the calculations to a small number of real space correlators, symmetries between real and spin space reduce the number of inequivalent ones further. In addition to the on-site and nearest neighbour correlations we only need to include the three third nearest neighbour correlations for non-zero $\Gamma$ capturing the leading longer range correlations in space.

\begin{figure}
	\centering
	\includegraphics[width=1.0\linewidth]{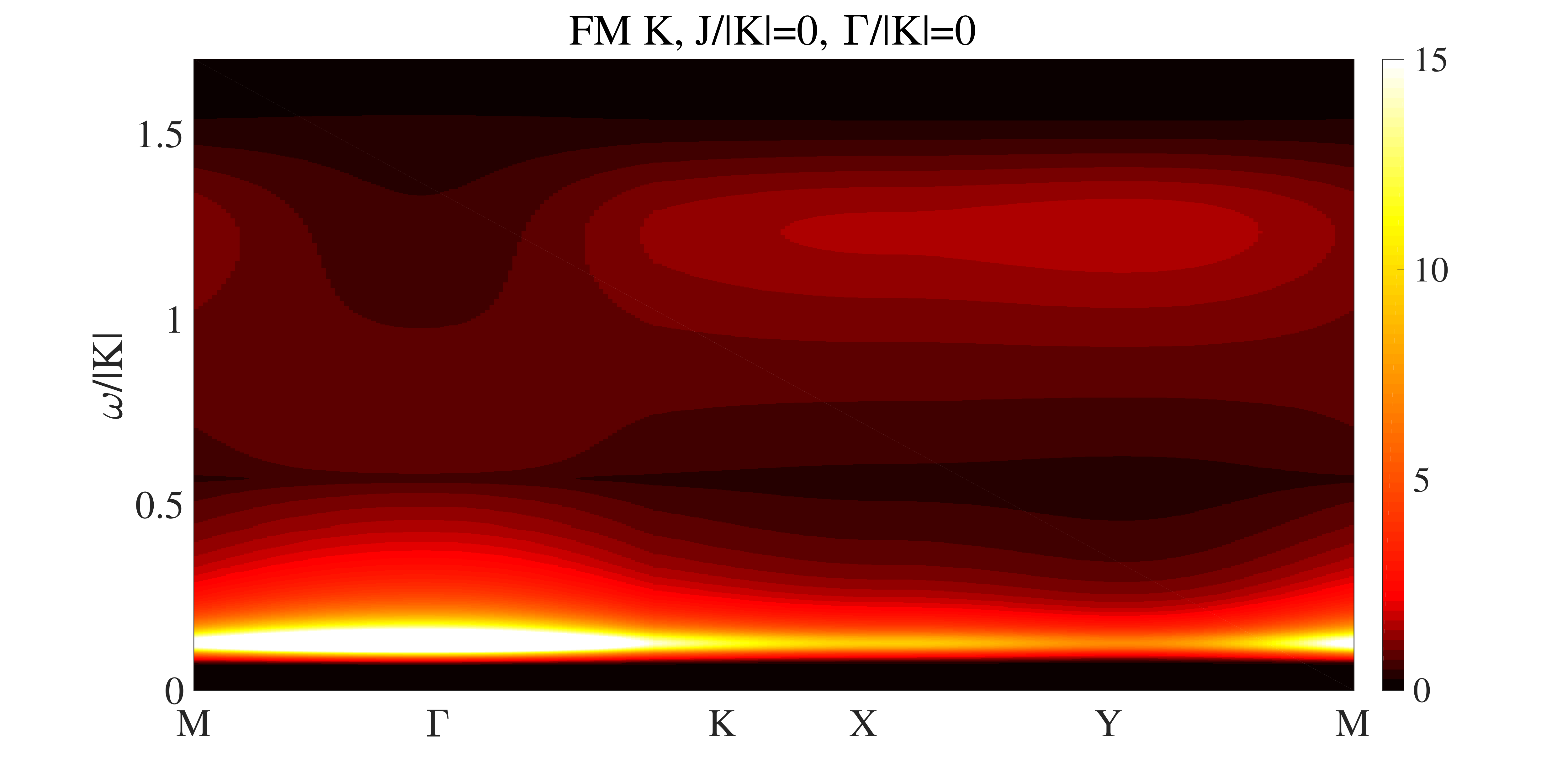}
	\includegraphics[width=1.0\linewidth]{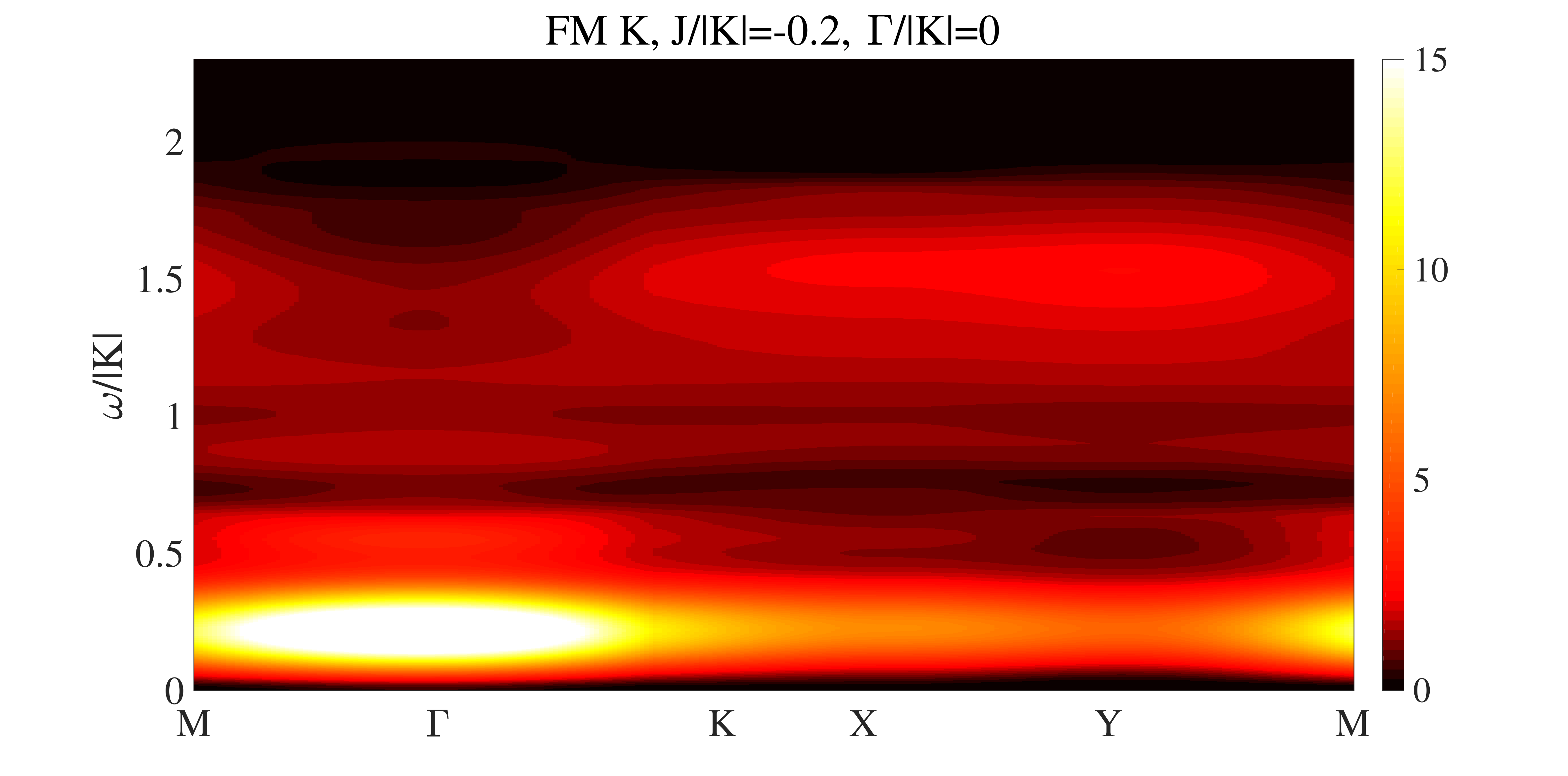}
	\includegraphics[width=1.0\linewidth]{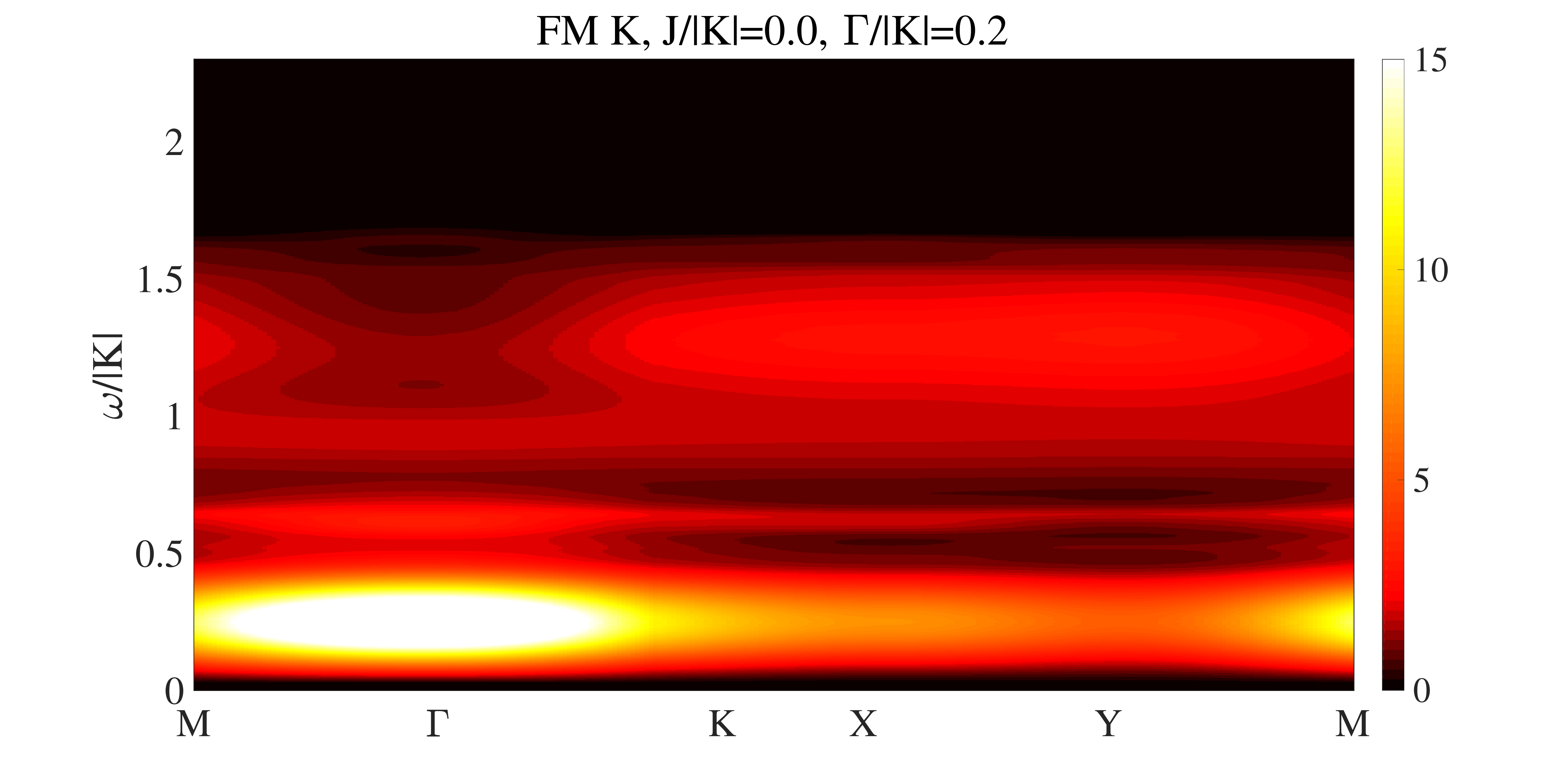}
	\includegraphics[width=1.0\linewidth]{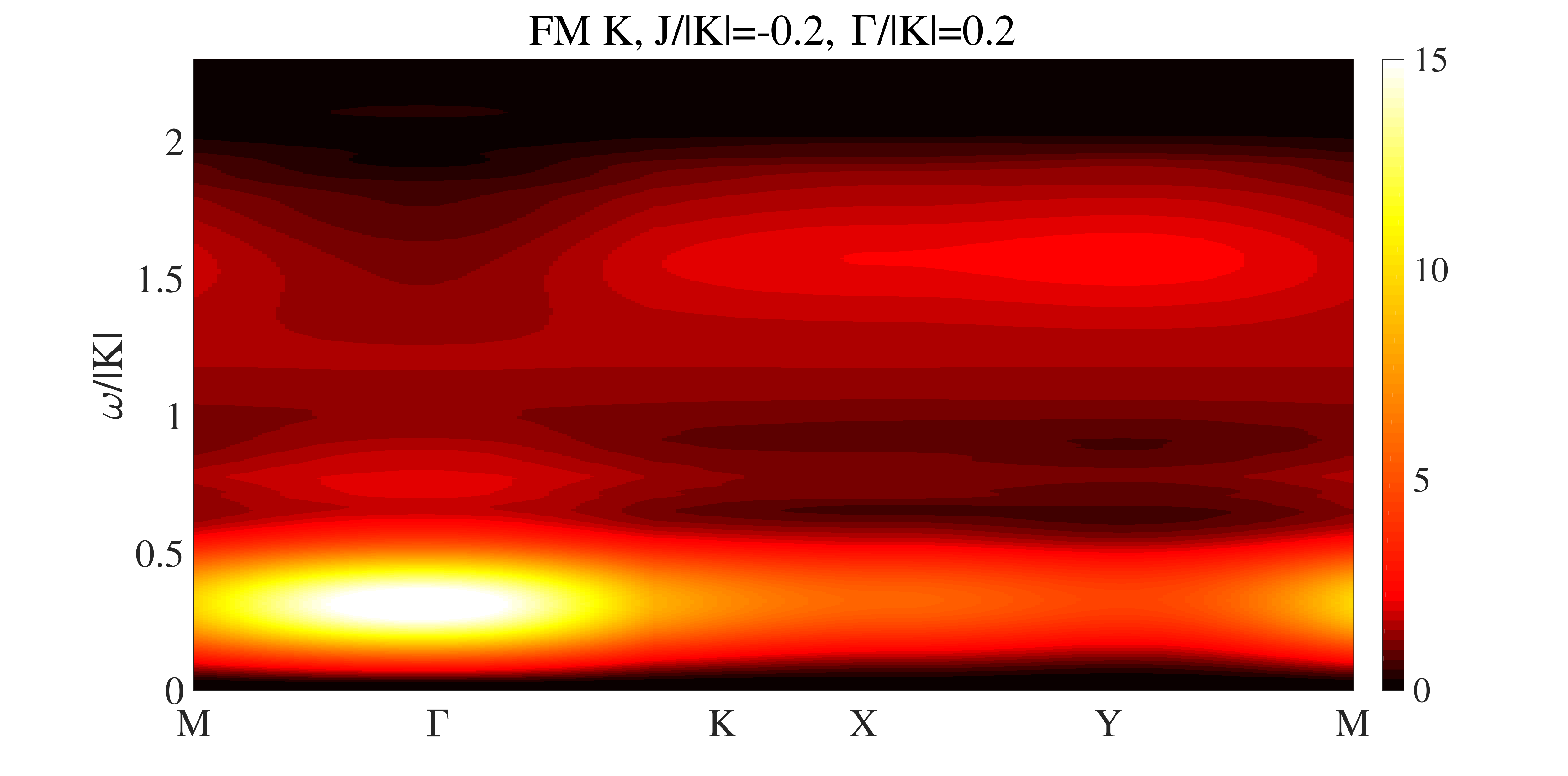}
		\caption{The dynamical structure factor $\mathcal{S}(\mathbf{q},\omega)$ is shown along a path through the BZ for different parameters of the Kitaev QSL phase.}
	\label{Sqw1}
\end{figure}
\begin{figure}
	\centering
	\includegraphics[width=1.0\linewidth]{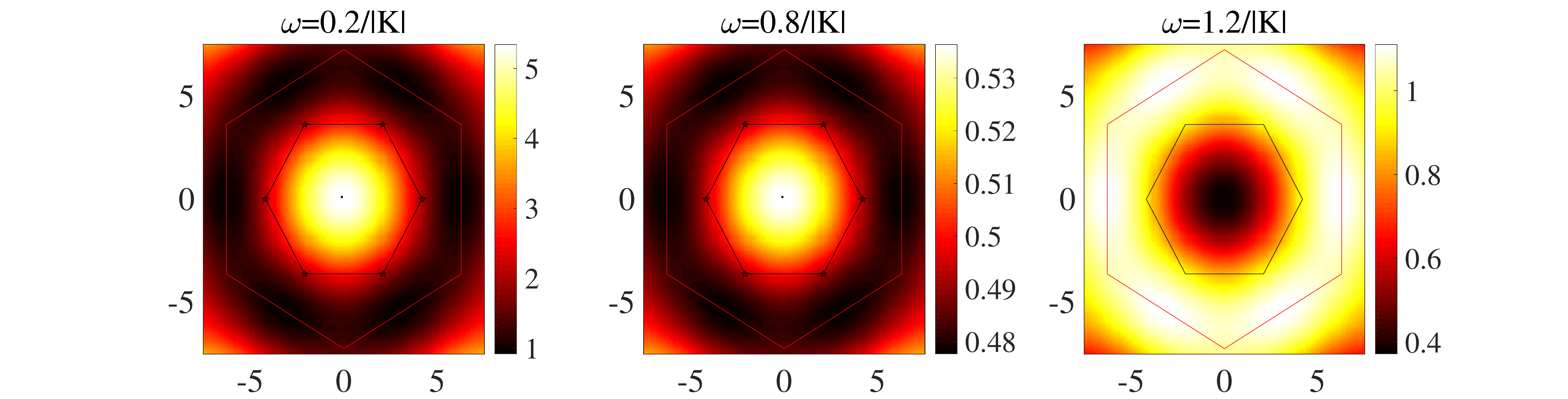}
	\includegraphics[width=1.0\linewidth]{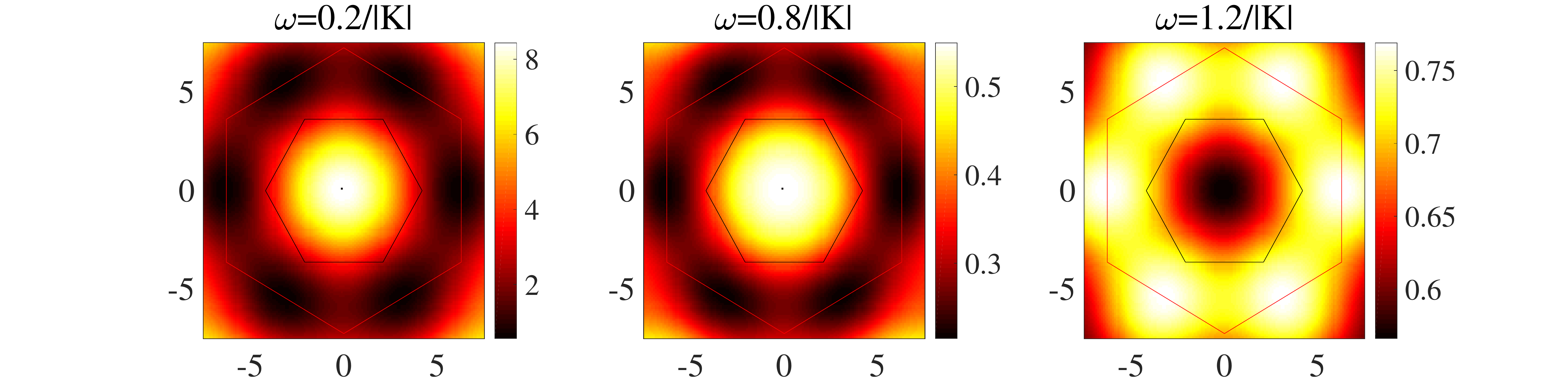}
	\includegraphics[width=1.0\linewidth]{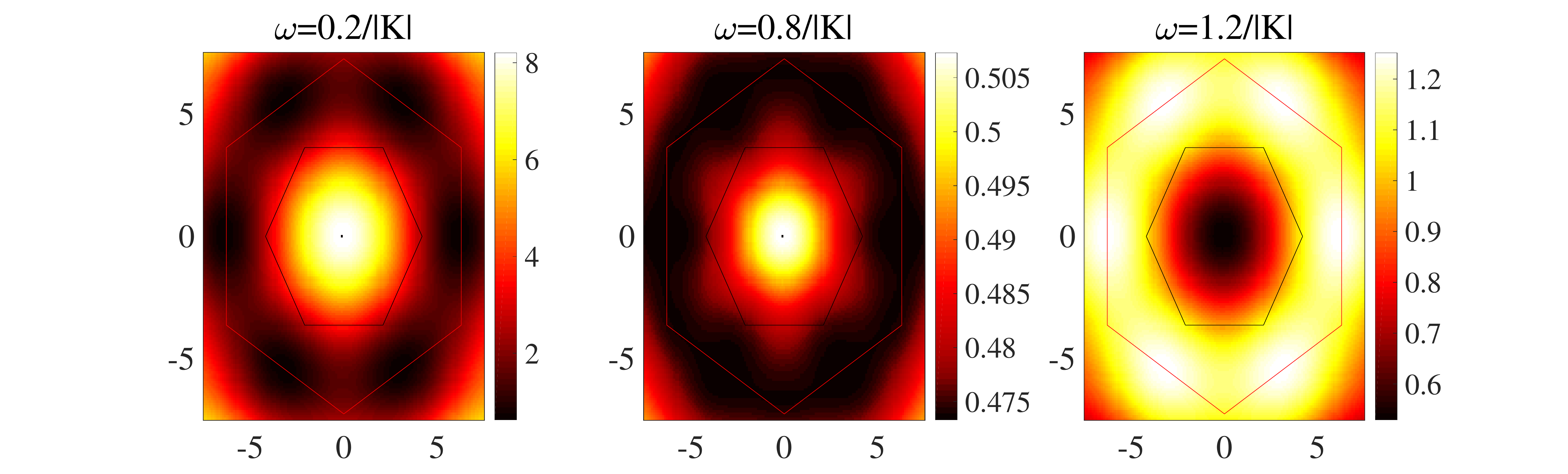}
	\includegraphics[width=1.0\linewidth]{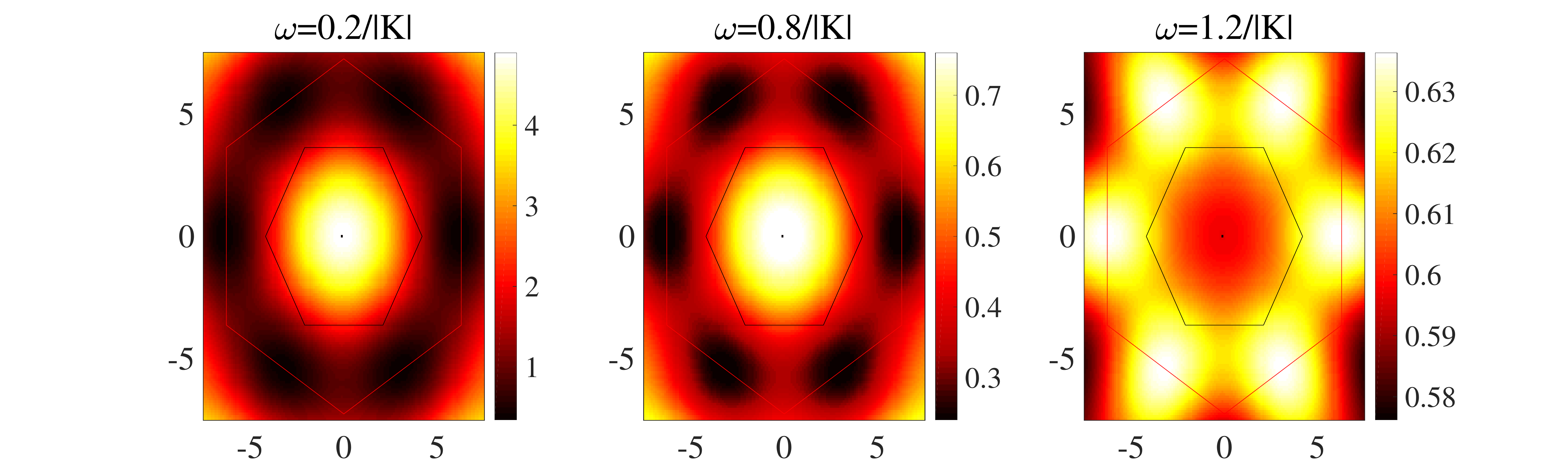}
		\caption{The dynamical structure factor $\mathcal{S}(\mathbf{q},\omega)$ is shown over the extended BZ for three different constant frequency cuts. The first (black) and second (red) BZs are indicated  The parameters of each panel are the same as for the corresponding panels in Fig.\ref{Sqw1}.}
	\label{Sqw2}
\end{figure}
\begin{figure}
	\centering
	\includegraphics[width=1.0\linewidth]{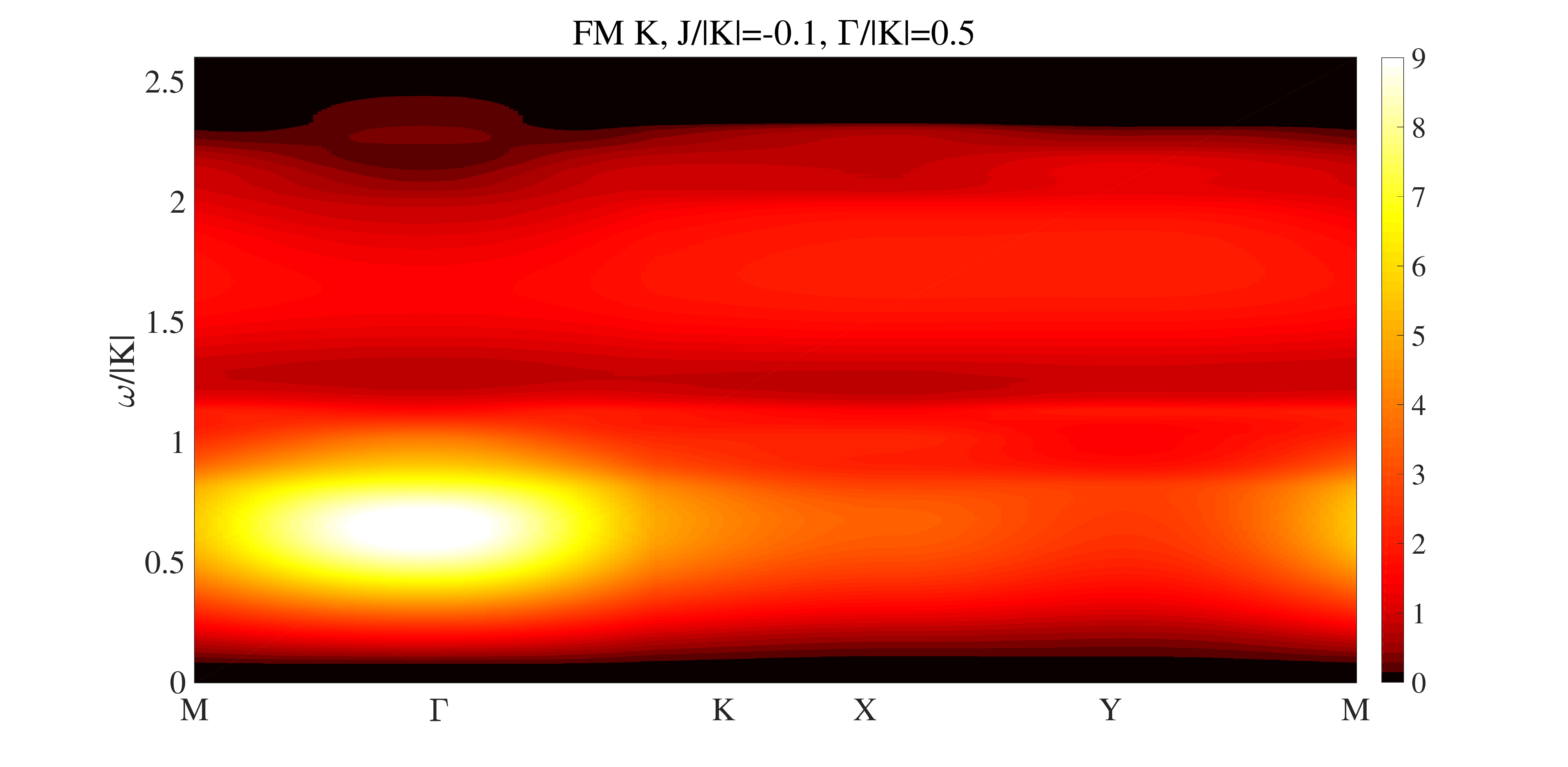}
	\includegraphics[width=1.0\linewidth]{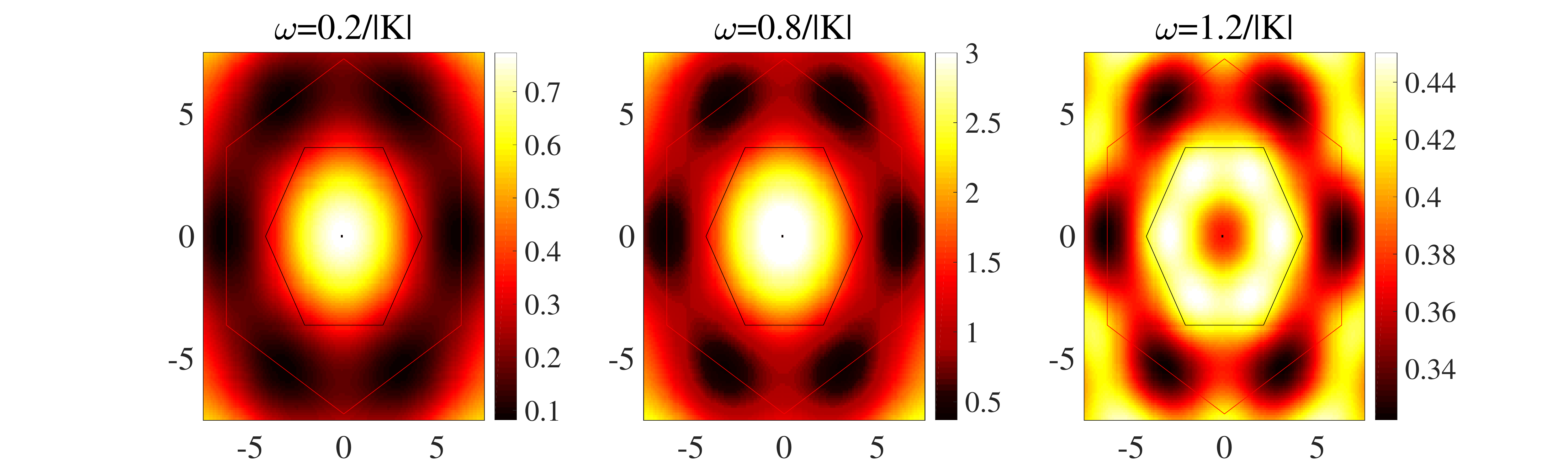}
	\includegraphics[width=1.0\linewidth]{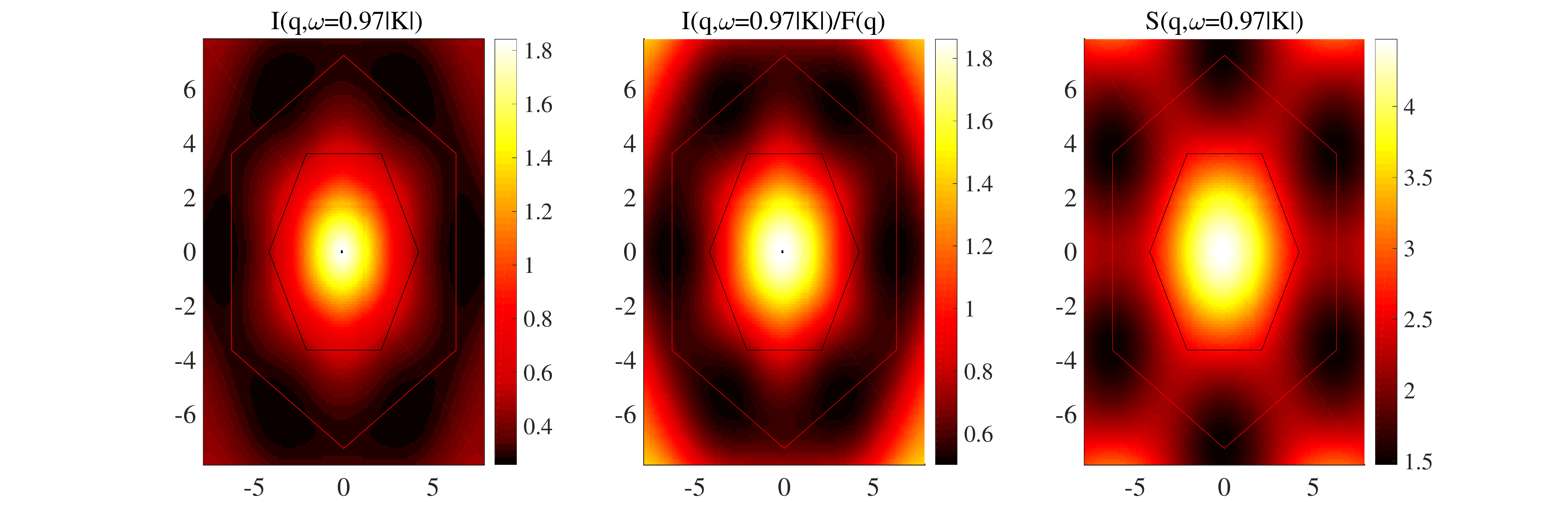}
		\caption{Dynamical spin correlation for the FM Kitaev QSL with the best-fit model parameters for $\alpha$-RuCl$_3$ from Winter {\it et al.}\cite{Winter2017} $J=-0.1$, $\Gamma=0.5$. Path through the BZ (upper panel) of the structure factor $\mathcal{S}(\mathbf{q},\omega)$ without the spin-momentum projectors (no form factor decay) and  constant frequency cuts (middle) including the projectors (no form factor decay). The lower panel shows the effect of the projectors for a constant frequency cut at $\omega=0.97 |K|$. The left panel shows the full INS scattering intensity $\mathcal{I}(\mathbf{q},\omega)$ , Eq.\ref{INSInt}, the middle panel without the decay $F(q)$ and the right panel the bare structure factor without the momentum projectors. It shows that the combination of the projector and the off-diagonal spin correlations induced by the $\Gamma$ term are one possible source for the star-like pattern observed in $\alpha$-RuCl$_3$.}
	\label{Sqw5}
\end{figure}
\begin{figure}
	\centering
	\includegraphics[width=1.0\linewidth]{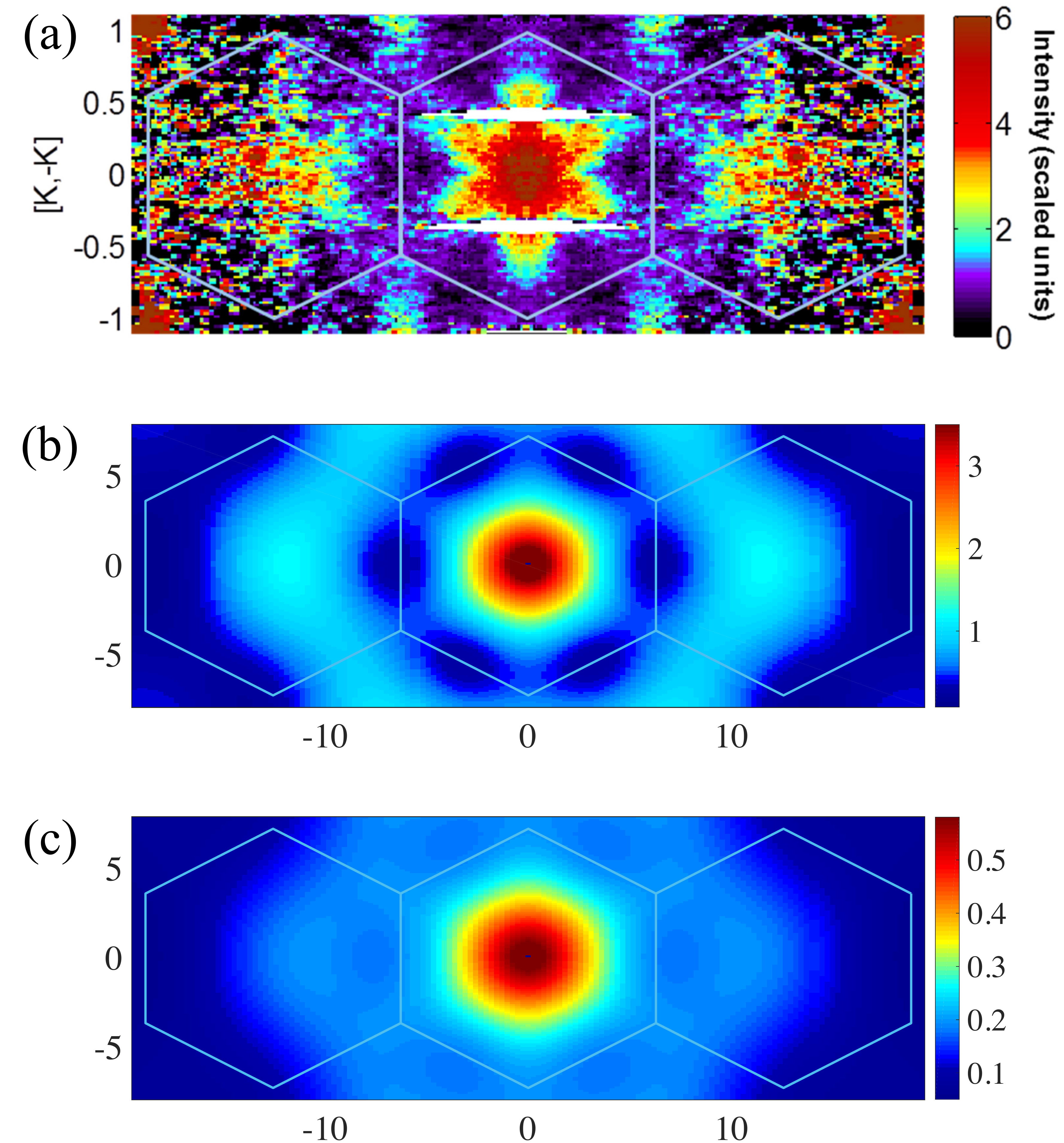}
		\caption{Comparison of the INS intensity for $\alpha$-RuCl$_3$ in panel (a), figure adapted from Ref.\onlinecite{Banerjee2017}, with that calculated for the best-fit model parameters for $\alpha$-RuCl$_3$ from Ref. Winter {\it et al.}\cite{Winter2017} $J=-0.1$, $\Gamma=0.5$ in panel (b) and  the exact solution of the pure Kitaev model panel (c). For concreteness we chose an energy $\omega/|K|=0.64$.}
	\label{Sqw8}
\end{figure}

\subsubsection{Dynamical correlations}
As has been pointed out previously, the gap in the structure factor of the 
pure Kitaev model is an artefact of the static fluxes of the integrable Kitaev 
limit~\cite{Song2016}. Similarly, we find that due to the fact that the suddenly inserted 
flux from a spin flip can hop away, away from the fine-tuned limit as captured by a finite life-time $\tau$, 
the gap of the structure factor is filled in. Hence, the overall phase $\Delta_b$ of 
Eq.\ref{SpinCorrSemiClassical} is adjusted self-consistently to obtain the 
gapless structure factor~\cite{Song2016} with a low frequency (long time) 
behaviour determined by the matter propagator $S(\omega)\propto G^c(\omega)\propto \omega$ 
(governed by the matter fermion DOS linear in $\omega$). Note, only at the Kitaev point do we have 
precisely $\Delta_b=-C^{\alpha}_{A0}$ but because of the factorization the overall phase is not simply given by one of the flux gaps.

Second, the new time scale $\tau$ has considerable effect on the  big low frequency peak dominating the dynamical
 structure factor of the pure Kitaev model, see Fig.~\ref{Sqzero2}. This is perhaps the visibly most striking consequence of integrability breaking -- more spectral weight is transferred 
to higher energies asymmetrically {\sbc{and anisotropically in presence of $\Gamma$}}. Moreover, the sharp peak at low energies is broadened in an asymmetric fashion and the the magnetic bandwidth is reduced/increased.

Third, the momentum dependence is changed because the strong bond selectivity is relaxed, 
e.g. a nonzero J induces $S^{xx/yy}$ correlations along a $z$-type bond, etc. In Figs.~\ref{Sqw1} and \ref{Sqw2} four representative regimes 
of the structure factor are shown. Note, however that the gross qualitative features of the FM Kitaev QSL -- a low energy peak in intensity at the $\Gamma$-point and a corresponding minimum of intensity for high energies -- is robust. Moreover, the overall magnetic bandwidth corresponding to the matter Majorana Fermions and the minimum of intensity at the energy scale of the van Hove singularity of the Majoranas fermions is another feature in common with the pure Kitaev model\cite{Knolle2014,Knolle2015}. The corresponding MFT band structures of the matter- and flux-Majorana Fermions 
are presented in Fig.~\ref{Energy_Dispersion}.

Finally, because of the dynamical nature of the fluxes, the intensity, which was mainly concentrated in a narrow low energy peak for the pure Kitaev point, is redistributed to higher energies.

\subsection{Connection to $\alpha$-RuCl$_3$}
Our work is partially motivated by the recent INS results on the honeycomb Kitaev candidate material  $\alpha$-RuCl$_3$ 
which shows unusual broad scattering continua reminiscent of the ones calculated for the pure Kitaev model~\cite{Banerjee2016,Banerjee2017,Do2017}. 
However, agreement is only partial, with  discrepancies concerning
 the overall intensity distribution as a function of frequency, as well as details of the momentum dependence. Note also that the low frequency response of $\alpha$-RuCl$_3$ at low temperatures should be related to the weak residual long range magnetic ordering. 
 
 Contributing to this ongoing discussion we show the  structure factor and INS intensity for the recently proposed best fit parameters~\cite{Winter2017} 
for $\alpha$-RuCl$_3$ in Fig.~\ref{Sqw5}. Note that despite the fact that the true ground state for these values 
for the exchange constants has weak long range magnetic order, our parton MFT continues to provide a (globally unstable) 
QSL solution. This is evidenced in that, for $\Gamma/|K|=0.5$ the flux gap goes to zero, see Fig.\ref{Energy_Dispersion} 
right panel. Therefore, strictly speaking this corresponds to a regime where our approximate treatment is 
uncontrolled. It is thus all the more striking that our present approach is able to give the following main 
qualitative changes {\sbc{raising the hope that in the actual material, the high energy spectral weight may remain unaffected by the low energy magnetic ordering and is determined by the proximate QSL physics.}}

In the frequency domain, the central feature is the dominant peak at low frequencies above the flux gap. This is much narrower 
in the integrable case than for our theory using the  best-fit parameters, which is indeed closer  to what is  found in experiment. 

In wavevector space, Fig.\ref{Sqw8}, the central peak for the integrable model is quite broad, covering the full Brioullin zone in that the minimum 
at the zone edge is relatively shallow (lower panel). This reflects the very short-range correlations of the integrable model. By contrast, 
for the best-fit model (middle panel), the corresponding minimum is considerably more pronounced, as it is in experiment (upper panel).

%The overall very broad scattering continuum consists of two different regimes separated in frequency. A low energy response dominated by a broad peak centred around the $\Gamma$-point and a higher energy response with more intensity around the boundaries of the BZ. 

Also in Fig.\ref{Sqw8}, a six-fold star pattern is very visible in the experimental data, which is almost entirely absent in the integrable case. 
A six-fold angular modulation is still rather weak in the best-fit model, but it is starting to be visible there, e.g.\ in the form of
higher scattering intensity at the $K$ points (Brioullin zone corners) compared to the $M$ points (midpoints of Brioullin zone  edges). This 
can be traced back to a combination of 
the momentum-spin projectors relating the structure factor to the INS intensity, see Eq.\ref{INSInt} and 
the spin off-diagonal terms induced by the $\Gamma$ term.% lead to a more pronounced 'star-like' pattern 
%of the momentum dependence in the constant frequency cuts which is reminiscent of the findings for $\alpha$-RuCl$_3$. 
We show a comparison of the scattering in the entire BZ with and without the momentum projectors in the lower panel of  Fig.\ref{Sqw5}, 
where the six-fold structure is considerably stronger.

Overall, it is encouraging that the best-fit parameters used for our augmented theory hence provide qualitative changes of the exact solution heading
in the correct direction with respect to experiment on $\alpha$-RuCl$_3$, but they still do not provide quantitative agreement. 
Whether this is due to the fact that the `perturbations' in the experimental system there are simply too large, or that the best-fit parameters
do not present the final answer themselves, or both, is something we are at present unable to settle. 

\section{Discussion and Outlook}
\label{sec:disc}

Beyond the immediate desire to account for experimental results in Kitaev candidate compounds, we would finally like to address 
how this work fits into the broader picture of studies of quantum spin liquids and fractionalised correlated phases more generally. 

It has been known for a long time that the ground state of the Kitaev honeycomb model can exactly be captured by a parton-based mean-field theory.
What was possibly less appreciated is that this does not extend to excited states; and by dint of this, to the dynamics even at zero temperature. Our theory
plugs this gap, and hence embeds the exactly soluble Kitaev model in the lore of parton theories, from the perspective of which a dynamical analysis
like the one possible here is non-trivial in itself. 

This exactly soluble augmented parton theory thus provides a controlled starting point for the inclusion of integrability-breaking terms. We choose to 
include the popular Heisenberg and $\Gamma$ terms, as these are believed to play a role in an experimentally relevant setting, as well as 
providing different flavours of integrability-breaking, both each in isolation, and in combination.  

Our treatment focuses on the leading effect of these perturbations on the `bulk' of the response, as loosely defined by spectral weight. Our 
treatment thus 
omits some terms which are perturbatively small (higher order) in the perturbations, such as a term of order $J_H^2\Gamma^2$, which however
is dominant in the long-distance limit, where it engenders an algebraically small tail to the correlations with a tiny prefactor. By contrast, our theory 
does account for a lower-order shift of the bulk of the spectral weight. In this sense, our analysis should be seen as a complement to that of Ref.\onlinecite{Song2016}.

Of course, when these high-order perturbations cease to be small, our theory ceases to be justified. It will thus, in its present simple form, systematically
fail to account with any degree of accuracy for the termination of the spin liquid phase, which would generically go along with a breakdown of the fractionalisation
in the form of a re-binding of the partons, and an increase (and eventually divergence) of the weight in the long-range part of the correlations. On top of this,
our \hflux\ approximation is not systematic, and there is presumably still considerable scope for alternative, optimised schemes beyond this first pass. Our derivation and first analysis of this augmented MFT immediately points in a number of directions for future work.

A first task is to extend the reach of the present analysis by targeting a larger class of experimental observables and computing
the relevant responses of other experimentally relevant probes, such as ESR\cite{Wang2017,Little2017}, Raman scattering~\cite{Sandilands2015,Glamazda2017,Nasu2016,Perreault2015,Perreault2016} or resonant inelastic X-ray scattering (RIXS)~\cite{Halasz2017}. 

Perhaps the most obvious new direction is the inclusion of other perturbations of 
experimental interest, most saliently a magnetic field\cite{Baek2017,Banerjee2017b,Sears2017}. This would allow to contrast the effect of the removal of time-reversal symmetry with and without
retention of integrability. 

Next on the wishlist would be the extension to other regimes;  a finite-temperature treatment\cite{Nasu2014} again being motivated by the
experimentally suggested existence of a proximate spin liquid whose 'parasitic' magnetic order melts above a finite N\'eel temperature, presenting
a finite-temperature magnetically disordered target regime\cite{Banerjee2016}.

On the technical/conceptual front, a more comprehensive inclusion of the spinon-vison (Majorana-flux) coupling is clearly desirable, as this would
extend the theory towards the inclusion of instabilities of the spin liquid phase. Such an analysis would also be of interest to other cases
of popular spin liquids, such as the perennial problem of the kagome lattice quantum antiferromagnet\cite{Norman2016}. 

Considerably more ambitious would be the extension of these ideas to other types of many-body states. A theory of a fractionalised magnetically
ordered phase would be a worthy goal indeed in this framework. 

Overall, we feel that our novel augmented $Z_2$ parton MFT strikes an attractive balance between solubility, controllability, and genericity. Hence, we believe it will be of more general interest for the study of topological QSLs beyond the example presented here of the paradigmatic Kitaev QSL phase.  

\section*{Acknowledgements}
We acknowledge helpful discussions with Yong Baek Kim, Frank Pollmann, Achim Rosch, Adam Smith and Roser Valenti. We are also very grateful
to Ganapathy Baskaran, John Chalker, Matthias Gohlke, Dima Kovrizhin, Yong Baek Kim, Saptarshi Mandal, Frank Pollmann, Robert Schaffer, Krishnendu Sengupta,  
R. Shankar and Ruben Verresen for collaboration on related subjects.   
 {{SB acknowledges MPG for funding through the partner group at ICTS, MPIPKS for hospitality and support 
 of the DST (India) project ECR/2017/000504.}} This work was in part supported by the Deutsche Forschungsgemeinschaft
 under grant SFB 1143.
 Statement of compliance with EPSRC policy framework on research data: All data accompanying this publication are directly available within the publication.

%%%%%%%%%%%%%%%%%

\bibliography{KitaevMFT_refs}

%%%%%%%%%%%%%

\appendix
\section{Mean-field solution and self-consistency equations}
\label{app:mfsolution}
%%%%%%%%%%%%%%%%%%%%%%%11111111111111111111111111111111111111111--moved to appendix
\subsection{MFT solutions $c$-type matter Majorana Fermions}
In the vicinity of the pure Kitaev QSL the ground state remains in the flux-free sector with all $u_{ij}=i b_i^{\alpha} b_j^{\alpha}=1=\sigma_{ij}^{\alpha}$. Here we show details of the diagonalization of the MF Hamiltonain Eq.\ref{MFTHamiltonian} in the ground state flux sector. Note that the additional interactions only change the overall bandwidth of the $c$-type Hamiltonian but leave the dispersion and wave functions unchanged.
First, we concentrate on the matter Majorana fermions. Introducing complex fermions with a new notation with the two sublattices labeled by $A/B$ 
\begin{eqnarray}
\label{complexFermionsC}
c_{A \mathbf{r}} & = & (f_{\mathbf{r}}+f_\mathbf{r}^{\dagger}) \    \   \text{and}   \  c_{B \mathbf{r}} =  i(f_{\mathbf{r}}-f_\mathbf{r}^{\dagger}) 
\end{eqnarray}
and their Fourier transformation $f_{\mathbf{r}}  =  \frac{1}{\sqrt{N_b}} \sum_{\mathbf{q}} e^{-\mathbf{q}\mathbf{r}} f_{\mathbf{q}}$
the Hamiltonian can be written in a standard Bogoliubov-de-Gennes form 
\begin{eqnarray}
\label{HamBdG_c}
H^c & = & \sum_{\mathbf{q}} 
\begin{bmatrix}
f^{\dagger}_{\mathbf{q}} & f_{-\mathbf{q}} 
\end{bmatrix}
\begin{bmatrix}
\xi_{\mathbf{q}} & -\Delta_{\mathbf{q}} \\
-\Delta_{\mathbf{q}}^* & -\xi_{\mathbf{q}}
\end{bmatrix}
\begin{bmatrix}
f_{\mathbf{q}} \\
f_{\mathbf{-q}}^{\dagger}
\end{bmatrix}
\end{eqnarray}
with the definitions
\begin{eqnarray}
\label{DefDispGap_c}
\!\!\!\!\!\!\!S(\mathbf{q}) \!\!&=& \!\!\left[ \frac{K+J}{4} \chi^b_K + \frac{J}{2} \chi^b_J +\frac{\Gamma}{2} \chi^b_{\Gamma} \right] \left( 1\!+\! e^{i\mathbf{q} \mathbf{n_x}} \!+\! e^{i\mathbf{q} \mathbf{n_y}}\right)
\end{eqnarray}
 $\xi_{\mathbf{q}}  =  \text{Re} S(\mathbf{q})$  and
$\Delta_{\mathbf{q}}  =  - i \text{Im} S(\mathbf{q})$.  The lattice vectors are $\mathbf{n_{x/y}}=(\pm\frac{1}{2},\frac{\sqrt{3}}{2})$ and the vector connecting the sublattices is $\mathbf{\delta}=(0,-\frac{1}{\sqrt{3}})$, see Fig.\ref{RealspaceCorrelations} (a).

A standard Bogoliubov rotation 
\begin{eqnarray}
\label{Bogoliubov_c}
\begin{bmatrix}
f_{\mathbf{q}} \\ f^{\dagger}_{-\mathbf{q}}
\end{bmatrix}
=
\begin{bmatrix}
\cos \theta_{\mathbf{q}} & i \sin \theta_{\mathbf{q}} \\
i \sin \theta_{\mathbf{q}} & \cos \theta_{\mathbf{q}}
\end{bmatrix}
\begin{bmatrix}
a_{\mathbf{q}} \\
a_{-\mathbf{q}}^{\dagger}
\end{bmatrix}
\end{eqnarray}
with $\tan 2\theta_{\mathbf{q}}= - \frac{\text{Im} S(\mathbf{q})}{\text{Re} S(\mathbf{q})}$ diagonalizes the system
\begin{eqnarray}
\label{DiagonalHamiltonian_c}
H^c=\sum_{\mathbf{q}} 2 |S(\mathbf{q})| \left[ a^{\dagger}_{\mathbf{q}} a_{\mathbf{q}} -\frac{1}{2} \right] 
\end{eqnarray}
such that the Heisenberg equation of motion directly gives the time dependence of the operators
\begin{eqnarray}
\label{TimeDepc}
a_{\mathbf{q}} (t) & = & a_{\mathbf{q}} e^{-i t 2 |S(\mathbf{q})|}.
\end{eqnarray}

%%%%%%%%%%
\subsection{MFT solutions $b$-type flux Majorana Fermions}
First, for $\Gamma=0$ the Hamiltonian of the different $b$-type Majoranas decouples and we can introduce three different complex fermions with $\alpha=x,y,z$ [and define $\mathbf{n_z}=(0,0)$]
\begin{eqnarray}
\label{complexFermionsB}
b^{\alpha}_{A \mathbf{r}} & = & (f^{\alpha}_{\mathbf{r}}+f_\mathbf{r}^{\alpha \dagger}) \    \   \text{and}   \  b^{\alpha}_{B \mathbf{r+n_{\alpha}}} =  i(f^{\alpha}_{\mathbf{r}}-f_\mathbf{r}^{\alpha\dagger}) 
\end{eqnarray}
and their Fourier transformation $f^{\alpha}_{\mathbf{r}}  =  \frac{1}{N_B} \sum_{\mathbf{q}} e^{-\mathbf{q}\mathbf{r}} f^{\alpha}_{\mathbf{q}}$.
The b-type MFT Hamiltonian can be written in a standard Bogoliubov-de-Gennes form 
\begin{eqnarray}
\label{HamBdG_b}
H^b & = & \sum_{\alpha} \sum_{\mathbf{q}} 
\begin{bmatrix}
f^{\alpha\dagger}_{\mathbf{q}} & f^{\alpha}_{-\mathbf{q}} 
\end{bmatrix}
\begin{bmatrix}
\xi^{\alpha}_{\mathbf{q}} & -\Delta^{\alpha}_{\mathbf{q}} \\
-\Delta^{\alpha*}_{\mathbf{q}} & -\xi^{\alpha}_{\mathbf{q}}
\end{bmatrix}
\begin{bmatrix}
f^{\alpha}_{\mathbf{q}} \\
f_{\mathbf{-q}}^{\alpha\dagger}
\end{bmatrix}
\end{eqnarray}
with the definitions
$\xi^{\alpha}_{\mathbf{q}}  =  \frac{K\chi^c}{4} +\frac{J \chi^c}{4}\text{Re} \sum_{\beta \neq \alpha} e^{-i \mathbf{q} \left( \mathbf{n_{\beta} -\mathbf{n_{\alpha}}}\right)}$ and $\Delta^{\alpha}_{\mathbf{q}}  =  i \frac{J \chi^c}{4} \text{Im} \sum_{\beta \neq \alpha} e^{-i \mathbf{q} \left( \mathbf{n_{\beta} -\mathbf{n_{\alpha}}}\right)}$.

As before, a standard Bogoliubov rotation 
\begin{eqnarray}
\label{Bogoliubov_b}
\begin{bmatrix}
f^{\alpha}_{\mathbf{q}} \\ f^{\alpha \dagger}_{-\mathbf{q}}
\end{bmatrix}
=
\begin{bmatrix}
\cos \theta^{\alpha}_{\mathbf{q}} & i \sin \theta^{\alpha}_{\mathbf{q}} \\
i \sin \theta^{\alpha}_{\mathbf{q}} & \cos \theta^{\alpha}_{\mathbf{q}}
\end{bmatrix}
\begin{bmatrix}
a^{\alpha}_{\mathbf{q}} \\
a_{-\mathbf{q}}^{\alpha\dagger}
\end{bmatrix}
\end{eqnarray}
with $\tan 2\theta^{\alpha}_{\mathbf{q}}= - \frac{i \Delta^{\alpha}_{\mathbf{q}}}{\xi^{\alpha}_{\mathbf{q}}}$  diagonalizes the system
\begin{eqnarray}
\label{DiagonalHamiltonian_b}
H^b=\sum_{\mathbf{q}} 2 |S^{\alpha}(\mathbf{q})| \left[ a^{\alpha\dagger}_{\mathbf{q}} a^{\alpha}_{\mathbf{q}} -\frac{1}{2} \right].
\end{eqnarray}
such that the dispersion $ |S^{\alpha}(\mathbf{q})| = \sqrt{|\xi^{\alpha}_{\mathbf{q}}|^2+|\Delta^{\alpha}_{\mathbf{q}}|^2}$ determines the time dependence similar to Eq.\ref{TimeDepc}.

Second, for nonzero $\Gamma$ the different sectors $\alpha=x,y,z$ of the flux-type Majoranas are all coupled. We use the standard Fourier transform $b_{A\mathbf{r}}^{\alpha}  =  \frac{1}{\sqrt{N_b}} \sum_{\mathbf{q}} e^{-\mathbf{q}\mathbf{r}} b_{A\mathbf{q}}^{\alpha}$ and because Majoranas are their own adjoints we get $b_{A\mathbf{q}}^{\alpha \dagger}=b_{A\mathbf{-q}}^{\alpha}$. Then we write the Hamiltonian as
\begin{widetext}
\begin{eqnarray}
\label{HbMft}
H^b & = &\sum_{\mathbf{q}} 
\begin{bmatrix}
b_{A\mathbf{q}}^{x} & b_{A\mathbf{q}}^{y} & b_{A\mathbf{q}}^{z} 
\end{bmatrix}
i \hat M_{\mathbf{q}}
\begin{bmatrix}
b_{B\mathbf{-q}}^{x} \\
b_{B\mathbf{-q}}^{y} \\
b_{B\mathbf{-q}}^{z}
\end{bmatrix} = 
\sum_{\mathbf{q},j=1,2,3}  i \epsilon^j_{\mathbf{q}} \phi^A_{j\mathbf{q}} \phi^B_{j\mathbf{-q}}   \  \   \    \   \text{with} \\
\hat M_{\mathbf{q}} & = & - \frac{\chi^c}{4}
\begin{bmatrix}
(J+K)e^{i \mathbf{q} \mathbf{n_x}} +J( e^{i \mathbf{q} \mathbf{n_y}}+1) &  \Gamma & \Gamma e^{i \mathbf{q} \mathbf{n_y}} \\
  \Gamma & (J+K)e^{i \mathbf{q} \mathbf{n_y}} +J( e^{i \mathbf{q} \mathbf{n_x}}+1) & \Gamma e^{i \mathbf{q} \mathbf{n_x}} \\
   \Gamma e^{i \mathbf{q} \mathbf{n_y}} &  \Gamma e^{i \mathbf{q} \mathbf{n_x}} & (J+K) +J( e^{i \mathbf{q} \mathbf{n_x}}+e^{i \mathbf{q} \mathbf{n_y}})
\end{bmatrix}
\end{eqnarray}
\end{widetext}
and the singular value decomposition $\hat M_{\mathbf{q}} = \hat U_{\mathbf{q}} \hat \Lambda_{\mathbf{q}} \hat V^{\dagger}_{\mathbf{q}}$ with the positive real diagonal matrix diag$\Lambda_{\mathbf{q}} = \left[\epsilon^1_{\mathbf{q}},\epsilon^2_{\mathbf{q}},\epsilon^3_{\mathbf{q}}\right]$ and
$b_{A\mathbf{q}}^{\alpha} = \sum_j U^\alpha_{j\mathbf{q}} \phi^A_{j \mathbf{q}}$ and $b_{B\mathbf{-q}}^{\alpha} = \sum_j V^\alpha_{j\mathbf{q}} \phi^B_{j \mathbf{-q}}$ with $U^\alpha_{j\mathbf{q}}$ the matrix elements of $\hat U_{\mathbf{q}}^{\dagger}$ ($\alpha$ labels the columns) and $V^\alpha_{j\mathbf{q}}$ the matrix elements of $\hat V_{\mathbf{q}}$ ($\alpha$ labels the rows). 
Note, from $\hat M_{\mathbf{-q}} =\hat M_{\mathbf{q}}^*$ 
we get $\hat \Lambda_{\mathbf{-q}} =\hat \Lambda_{\mathbf{q}}$ and for the unitary matrices $\hat U_{\mathbf{-q}} =U_{\mathbf{q}}^*$ and $\hat V_{\mathbf{-q}} =V_{\mathbf{q}}^*$.

Finally, with the use of standard complex fermions
\begin{eqnarray}
\label{DefcomplexFermions}
\phi^A_{j\mathbf{q}} & = & f^b_{j\mathbf{q}} + f^{b\dagger}_{j\mathbf{-q}} \\
\phi^B_{j\mathbf{q}} & =  &-i \left\lbrace f^b_{j\mathbf{q}} - f^{b\dagger}_{j\mathbf{-q}} \right\rbrace \\
\end{eqnarray}
 we obtain 
\begin{eqnarray}
\label{BFermionsDiagonal}
H^b & = & \sum_{\mathbf{q},j=1,2,3} 2 \epsilon^j_{\mathbf{q}} \left\lbrace f^{b\dagger}_{j\mathbf{q}} f^b_{j\mathbf{q}} -\frac{1}{2}\right\rbrace.
\end{eqnarray} 
such that the dispersions $2 \epsilon^j_{\mathbf{q}}$ determines the time dependence similar to Eq.\ref{TimeDepc}.

\subsection{Self-consistency equations}
We have four different mean-field parameters each of which is determined by its own self-consistency equation. 
First, the one from the matter fermions
\begin{eqnarray}
\label{SelfConstEqns_c}
\chi^c & = & i \langle c_{A0} c_{B0}\rangle = 2 \sum_{\mathbf{q}} \cos^2 \theta_{\mathbf{q}} -1 = 0.5249 
\end{eqnarray}
which is independent of the values $K,J,\Gamma$ and $\chi^b_{J/K/\Gamma}$. 
There are three equations for the flux-fermions. 
\begin{eqnarray}
\label{SelfConstEqns_bK}
\chi^b_K & = & i \langle b^z_{A0} b^z_{B0}\rangle = - \sum_{\mathbf{q},j} U^z_{j\mathbf{q}} V^z_{j\mathbf{q}}  \\
\chi^b_J & = & i \langle b^x_{A0} b^x_{B0}\rangle = - \sum_{\mathbf{q},j} U^x_{j\mathbf{q}} V^x_{j\mathbf{q}} \\
\chi^b_{\Gamma} & = & i \langle b^x_{A0} b^y_{B0}\rangle = - \sum_{\mathbf{q},j} U^x_{j\mathbf{q}} V^y_{j\mathbf{q}}    
\end{eqnarray}

\begin{figure}
	\centering
	\includegraphics[width=1.1\linewidth]{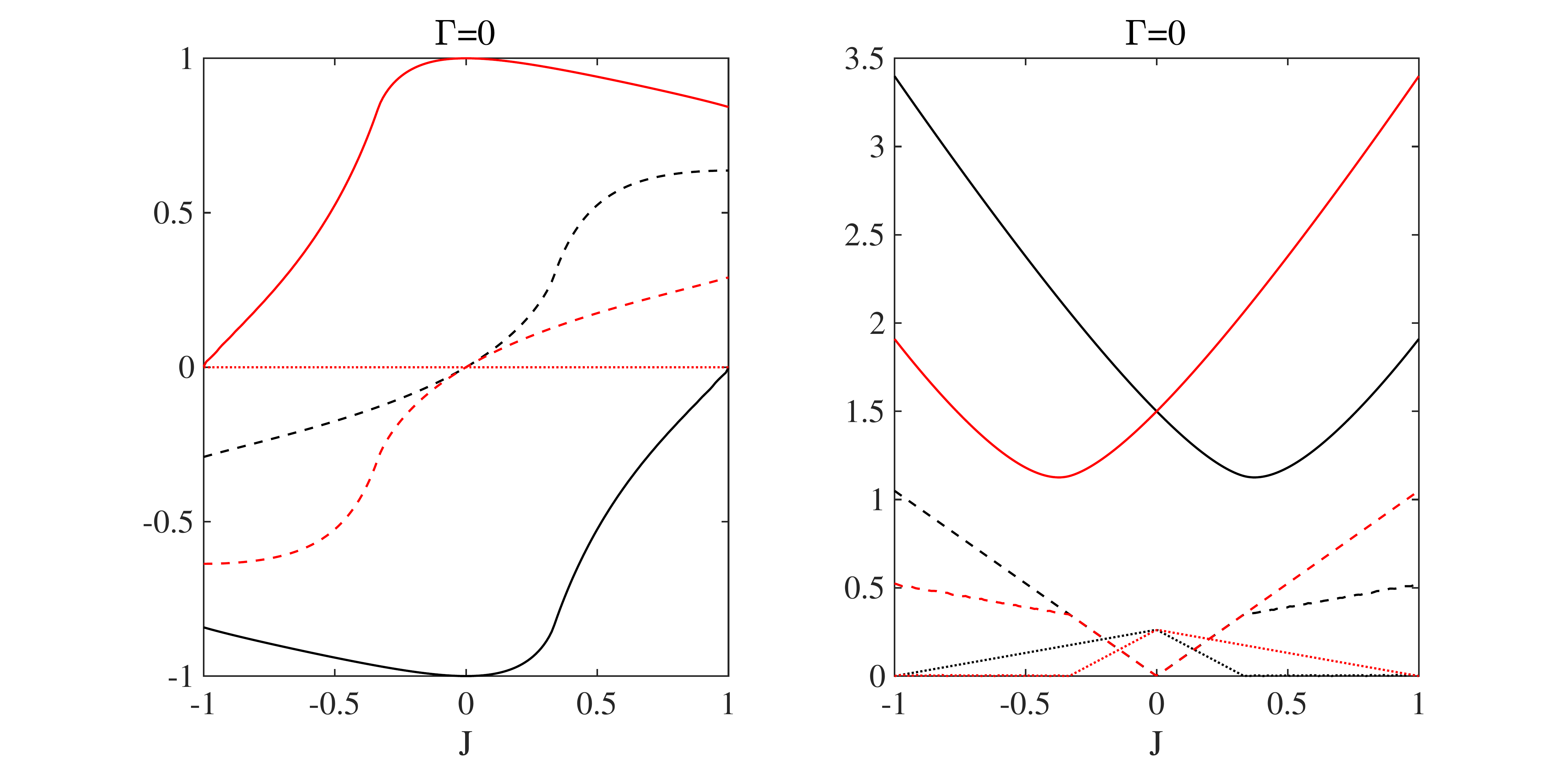}
	\includegraphics[width=1.1\linewidth]{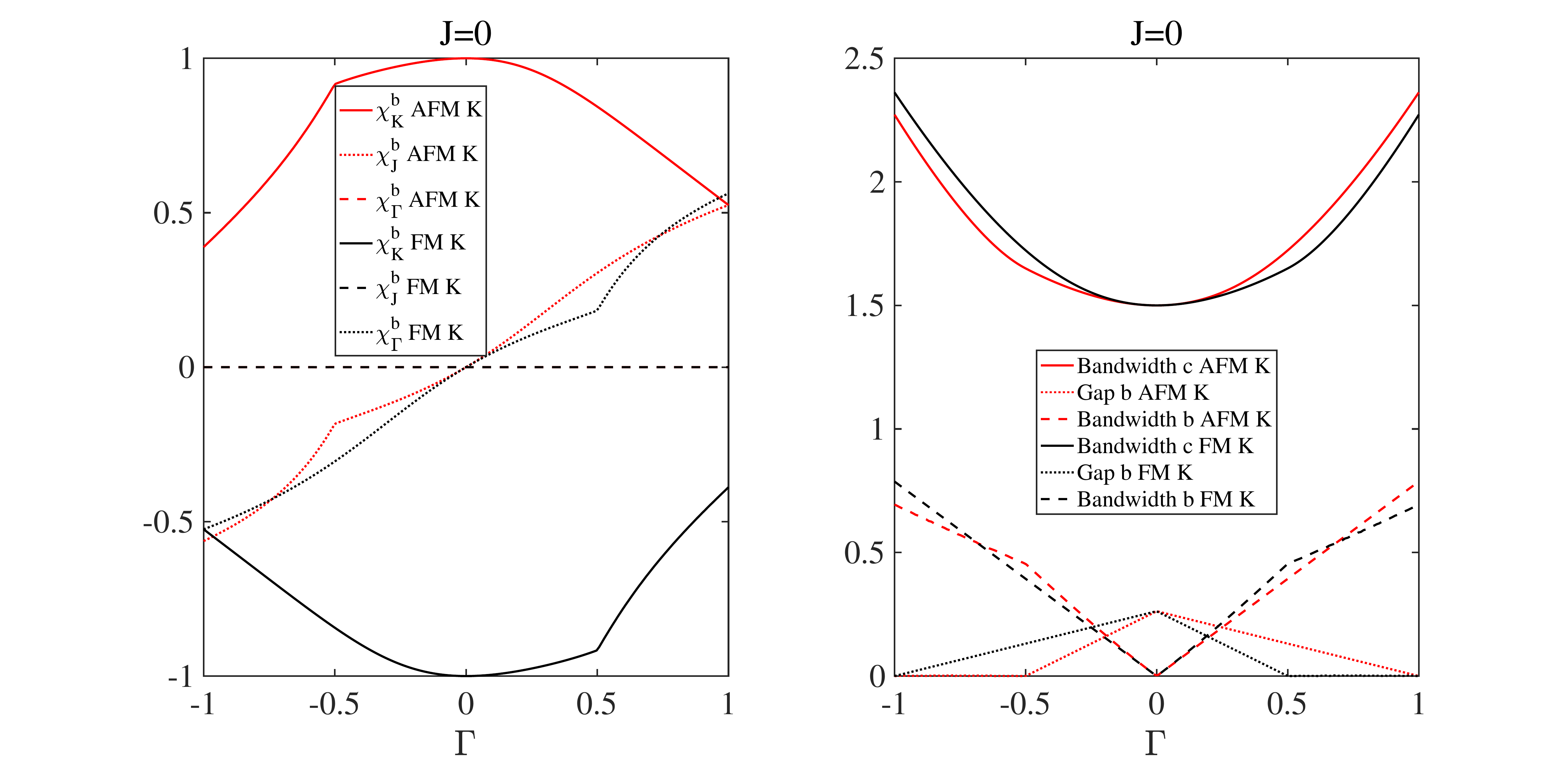}
	\includegraphics[width=1.1\linewidth]{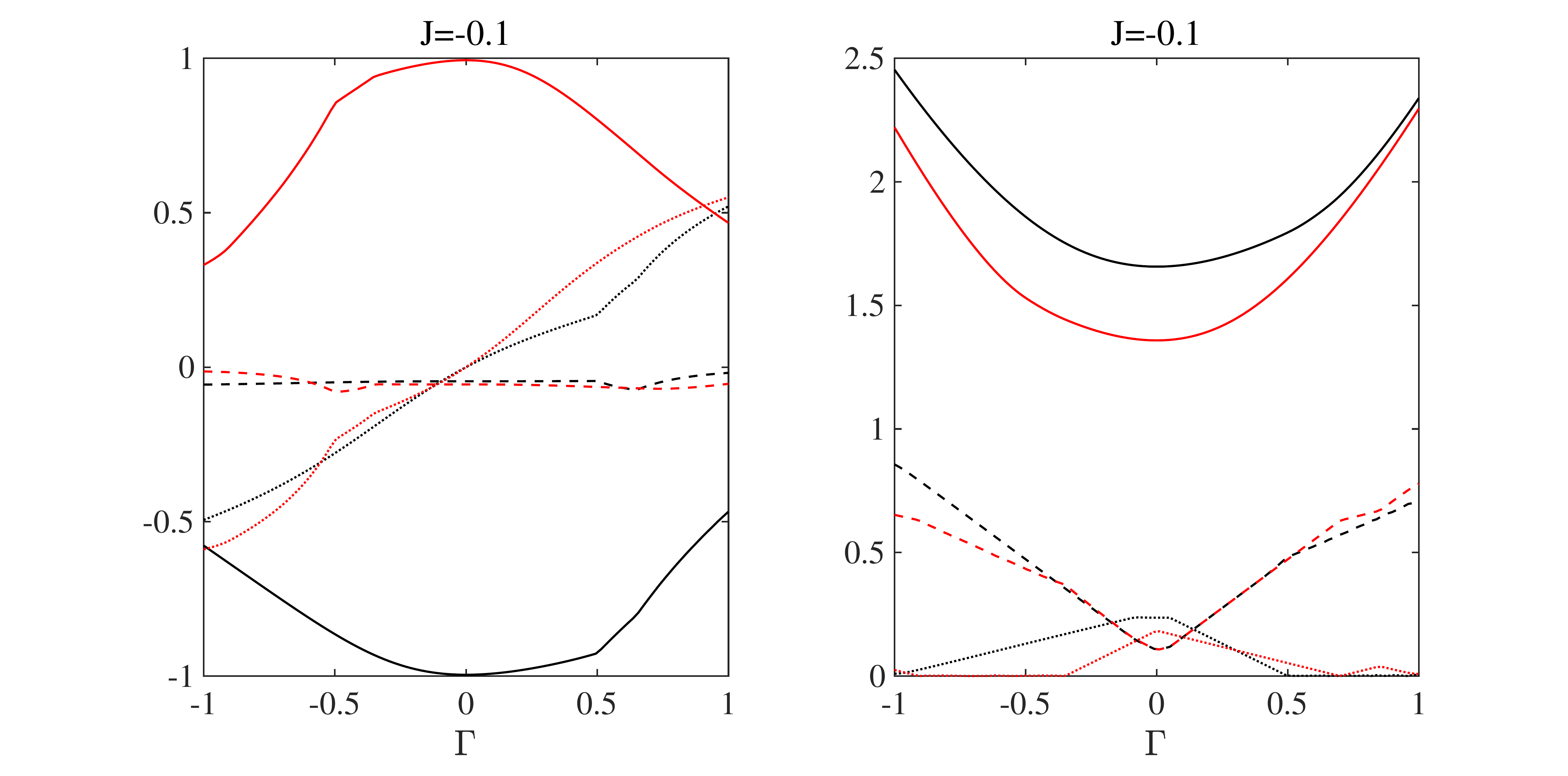}
		\caption{Evolution of the order parameters as a function of the exchange couplings (left panels); e.g. varying $J$ for $\Gamma=0$ (upper panels), varying $\Gamma$ for $J=0$ (middle panels), and varying $\Gamma$ for FM $J=-0.1$ (lower panels). The right panel shows the corresponding change in the bandwith for the $c$-type and $b$-type Majorana fermions and the evolution of the gap of the $b$'s, see the inset of the middle panels for the legends.}
	\label{Fig1}
\end{figure}

The evolution of the the order parameters as a function of the different exchange couplings is plotted in Fig.\ref{Fig1}. 
The MFT dispersions of the different types of excitations are shown in Fig.\ref{Energy_Dispersion} for representative parameter values. 

\end{document}